\documentclass[prx,twocolumn,aps,showpacs,amsmath,amssymb,superscriptaddress,floatfix,longbibliography]{revtex4-1}

\usepackage{color}
\usepackage{bm}
\usepackage{graphicx}
\usepackage{braket}
\newcommand{\bS}{{\bf S}}

\newcommand{\tp}{\tilde{\psi}}
\newcommand{\btp}{\tilde{\bm{\psi}}}

\usepackage[pdftex]{hyperref}
\hypersetup{colorlinks = true, urlcolor = blue, linkcolor = blue, citecolor = blue}

\begin{document}

\title{Strong coupling phases of the spin-orbit-coupled spin-1 Bose-Hubbard chain: odd integer Mott lobes and helical magnetic phases}
\author{J. H. Pixley}
\affiliation{Condensed Matter Theory Center and Joint Quantum Institute, Department of Physics, University of Maryland, College Park, Maryland 20742-4111 USA}
\affiliation{Department of Physics and Astronomy, Center for Materials Theory, Rutgers University, Piscataway, NJ 08854 USA}
\author{William S. Cole}
\affiliation{Condensed Matter Theory Center and Joint Quantum Institute, Department of Physics, University of Maryland, College Park, Maryland 20742-4111 USA}
\author{I. B. Spielman}
\affiliation{Joint Quantum Institute, National Institute of Standards and Technology, and University of Maryland, Gaithersburg, Maryland, 20899, USA}
\author{Matteo Rizzi}
\affiliation{Universit{\" a}t Mainz, Institut f{\" u}r Physik, Staudingerweg 7, D-55099 Mainz, Germany}
\author{S. Das Sarma}
\affiliation{Condensed Matter Theory Center and Joint Quantum Institute, Department of Physics, University of Maryland, College Park, Maryland 20742-4111 USA}

\date{\today}

\begin{abstract}
We study the odd integer filled Mott phases of a spin-1 Bose-Hubbard chain and determine their fate in the presence of a Raman induced spin-orbit coupling which has been achieved in ultracold atomic gases;
this system is described by a quantum spin-1 chain with a spiral magnetic field.
The spiral magnetic field initially induces helical order with either ferromagnetic or dimer order parameters, giving rise to a spiral paramagnet at large field.
The spiral ferromagnet-to-paramagnet phase transition is in a novel universality class, with critical exponents
associated with the divergence of the correlation length $\nu \approx 2/3$ and the order parameter susceptibility $\gamma \approx 1/2$. 
We solve the effective spin model exactly using the density matrix renormalization group, and compare with both a large-$S$ classical solution and a phenomenological Landau theory. 
We discuss how these exotic bosonic magnetic phases can be produced and probed in ultracold atomic experiments in optical lattices.
\end{abstract}
\date{\today}

\maketitle

Strongly-correlated  quantum spin chains are an interacting many-body system that have been an instrumental platform to develop an understanding of topological properties~\cite{Pollman-2012}, Berry phase effects~\cite{Haldane-1983}, and quantum phase transitions~\cite{Affleck-1987}. One of the paradigmatic theoretical models in this context is the spin-1 bilinear-biquadratic Heisenberg chain~\cite{Haldane-1983,Affleck-1987,AKLT-1987} 
\begin{equation}
H_{BQ}/J = \sum_i \cos{\theta} ({\bf S}_i\cdot{\bf S}_{i+1})+\sin\theta ({\bf S}_i\cdot{\bf S}_{i+1})^2,
\label{eqn:ham0}
\end{equation}
($J$ is the unit of energy) which supports several conceptually important phases and phase transitions~\cite{Uimin-1970,Lai-1974,Sutherland-1975,Takhtajan-1982,Babujian-1982,Klumper-1989,Sorensen-1990, White-1993, Matteo-2005, Lauchli-2006}. 
The physics associated with the Hamiltonian in Eq.~(\ref{eqn:ham0}) is not only theoretically tantalizing, but is also directly accessible to experiments. For example, the antiferromagnetic  spin-1 chain ($\cos\theta>0$) has been probed experimentally in insulating quantum magnets~\cite{Ma-1992, Hagiwara-1990, Kenzelmann-2002}
that possess well-isolated, quasi-one-dimensional chains of $S=1$ local moments. Interestingly, ultracold atomic gases are  
a natural platform for realizing the complementary ferromagnetic spin-1 chain ($\cos\theta<0$), where the system is an effective description of the Mott insulating spin-1 Bose-Hubbard model~\cite{Imambekov-2003,Kimura-2005,Pai-2008,Forges-2013,Natu-2015,2013_SKurn_Ueda_RMP}. 
Solid state experiments always contain at least some disorder that breaks the spin chain into segments~\cite{Hagiwara-1990}, while the cold atomic gas setting is essentially pristine with no disorder in principle, so that the theoretical model defined by Eq.~(\ref{eqn:ham0}) (as well as its generalizations) can be studied directly.
Until recently, it has been difficult to realize magnetic ordering in ultra-cold atomic gases~\cite{Hart-2015,Mazurenko-2016}, and
a crucial element of physics is developing a theoretical understanding of the available strongly correlated phases.

A virtue of the flexibility and the precise control in atomic and molecular experiments is that various physical effects can be engineered simulating desired features of solid-state systems despite the vastly different setting (i.e.~atoms versus solids). 
Along these lines, recent experiments on cold atomic gases have demonstrated that, despite the atoms being electrically neutral, spin orbit coupling (SOC) can be engineered using two counter propagating Raman lasers in gases of bosons~\cite{Lin-2009,Lin-2011} or fermions~\cite{Wang-2012}.  For spin-1 bosons  using the setup at the National Institute of Standards and Technology~\cite{Campbell-2016}, this causes the single particle dispersion to have three degenerate minima, allowing the bosons to condense at three distinct values of quasimomentum~\cite{Ueda-12,Cole2015,Lan2014}. In an optical lattice this system supports an itinerant (i.e. superfluid) spin density wave phase that is suppressed as the strength of interactions is increased~\cite{Pixley-2016}, and as a result the phase diagram becomes quite distinct from the case with no SOC.
In the presence of SOC the conventional magnetic orders describing the strong-coupling Mott limit no longer apply. For example, in the case of bosons with a pseudospin-$1/2$, a wide array of interesting magnetic phases have been studied that result from anisotropic (i.e., compass model) and Dzyaloshinskii-Moriya interactions that are characteristic of SOC~\cite{mott_2d_cole, mott_2d_radic, mott_2d_cai, mott_2d_gong, Hickey2014, mott_xu, mott_zhao1, mott_zhao2, mott_piraud, mott_peotta}. 
It is also possible that  quantum phase transitions in these systems could allow access to unusual, or even undiscovered, universality classes due to the interplay of magnetic interactions and spin orbit coupling.
Such interacting one-dimensional spin models with SOC are difficult to realize in solid state materials, thus making ultracold atomic systems unique in their potential for studying new exotic quantum phases in the laboratory which may not exist in solids at all.

For the spin-1 case (which in the absence of a SOC has an inherent SU(2) symmetry and is described by Eq.~(\ref{eqn:ham0}) for the odd-integer Mott lobes), essentially nothing is known about its insulating magnetic ground states in the presence of SOC (implemented as a helical Zeeman field). The present manuscript fills this gap and addresses the important question of the relevant quantum phase diagram of this strongly correlated SOC-coupled interacting magnetic system. We focus on the physics deep in the Mott phase of one-dimensional spin-1 bosons in the presence of SOC with an odd-integer filling, where charge excitations are gapped out, and the effective Hamiltonian is purely magnetic (i.e.~we can deal with just spins).
As we have already pointed out, in cold atom systems achieving the low spin-entropy necessary for realizing this quantum-magnetic model has proven challenging. Therefore, we propose a different experimental route.  It is straightforward to create initial atomic states with perfect spin polarization (i.e. with zero entropy).  In the present case of a spiral ferromagnet (as we will show), it is not difficult experimentally to initialize such a low entropy state and either adiabatically or diabatically move the system into a corresponding zero-spin entropy magnetically ordered state. Again, such a protocol essentially impossible to achieve in solid state materials.

Specifically, we first derive an effective Hamiltonian in the Mott insulating limit, which is a ferromagnetic bilinear-biquadratic spin-1 chain in the presence of a \emph{spiral magnetic field}. We then solve this effective model using the density matrix renormalization group (DMRG) to map out the zero temperature phase diagram in both the magnetic field (i.e., SOC) strength and the biquadratic interaction. We show how the phases that exist in the ferromagnetic model (ferromagnet and dimer phases) evolve into a spiral ferromagnet and a spiral dimer phase respectively. To obtain a better qualitative understanding of the phase diagram and the role of quantum fluctuations, we also consider the large-$S$ classical limit of the model, as well as a phenomenological Landau theory treatment of the spiral ferromagnet. We use finite size scaling of the ground state correlation functions to determine the static critical properties of the spiral ferromagnet-to-paramagnet quantum phase transition. We find that the correlation exponent $\nu\approx 2/3$ and the susceptibility exponent $\gamma \approx 1/2$ fall outside of any known universality class (to the best of our knowledge). Our work thus opens up the possibility of exploring novel quantum critical phenomena in ultra cold gases. As an aside, we point out that for the one-dimensional interacting spin system of our interest here, DMRG is essentially an exact, albeit numerical, technique for obtaining the ground state quantum phase diagram and the critical exponents underlying the corresponding quantum phase transitions.

The rest of the manuscript is organized as follows: In section~\ref{sec:model} we derive the effective spin model, which we proceed to study with DMRG. In section~\ref{sec:sf} we focus on the spiral ferromagnetic phase, and in section~\ref{sec:sd} we study the spiral dimer phase. We bring these results together to construct the full phase diagram in section~\ref{sec:pd}. We close with a discussion of our results and their implications for future cold atom experiments in section~\ref{sec:dc}.

\section{Model}
\label{sec:model}
Our starting point is the Bose-Hubbard model for spin-1 bosons in the presence of  (Raman induced) spin-orbit coupling, which is implemented physically in the lab frame as a spiraling Zeeman field~\cite{Higbie-2002,Lin-2011,Campbell-2016},

\begin{eqnarray}
H_{\mathrm{BH}} &=&
-t\sum_{ i,\alpha }\left(b_{i\alpha}^{\dag}b_{i+1\alpha}+\mathrm{H.c} \right)
+ \sum_i {\bf h}_i\cdot {\bf S}_i \nonumber \\
&+& \frac{U_0}{2}\sum_i n_i(n_i-1)  + \frac{U_2}{2}\sum_i \left(\bS_i^2-2n_i \right)
\label{eqn:bh_rotate}
\\
&\equiv& H_t + H_{h} + H_U.
\end{eqnarray}
The spin-independent hopping amplitude is $t$, and we include onsite density and spin interactions $U_0$ and $U_2$ respectively.
We consider chains of length $L$,
$b_{i\alpha}^{\dag}$ creates a boson at site $i$ with a spin $\alpha$, and we have introduced density $n_{i} = \sum_{\alpha} b_{i\alpha}^{\dag}b_{i\alpha}$ and spin ${\bf S}_{i} = \sum_{\alpha,\beta} b_{i\alpha}^{\dag}{\bf T}_{\alpha\beta}b_{i\beta}$ operators, where ${\bf T}$ denotes the vector of spin-1 angular momentum matrices. The interaction parameters $U_0$ and $U_2$ are individually determined by the $s$-wave scattering length and the lattice geometry, but the ratio $U_2/U_0 = (a_2-a_0)/(a_0+2a_2)$ is dictated only by the scattering length $a_S$ in the total spin $S$ sector~\cite{Imambekov-2003}. Finally, the magnetic field
\begin{equation}
{\bf h}_i=h \left( \cos(\eta r_i)\hat{x}  -\sin(\eta r_i)\hat{y}\right)
\end{equation}
implementing the SOC is characterized by two parameters: intensity $h$ and pitch $\eta$. 

To put this in the more familiar translation-invariant form describing SOC in solid-state systems, we can transform our spin basis to locally follow the external field. We can define a new set of appropriately rotated bosons,
\begin{equation}
a_{i \alpha} = \sum_{\beta}(e^{-i r_i \eta T_z})_{\alpha \beta}b_{i \beta},
\end{equation}
and the hamiltonian now reads
\begin{eqnarray}
H_{\mathrm{BH}} &=& -t\sum_{ i,\alpha,\beta }\left(a_{i\alpha}^{\dag}(e^{i \eta T_z})_{\alpha\beta}a_{i+1\beta}+\mathrm{H.c} \right)+ h\sum_i S_i^x
\nonumber
\\
&+&\frac{U_0}{2}\sum_i n_i(n_i-1) + \frac{U_2}{2}\sum_i \left(\bS_i^2-2n_i \right).
\label{eqn:bh}
\end{eqnarray}
The SOC here appears as a more conventional uniform matrix-valued Peierls' phase. Also, due to the spin dependent hopping in Eq.~(\ref{eqn:bh}) one can equivalently interpret the spin states ($\alpha$) as a ``synthetic dimension" with a lattice length $2S+1$ sites. In this geometry all of the interactions in the Bose-Hubbard model are  ``non-local'' (i.e. along the transverse synthetic `spin' dimension) and the phase $\eta$ due to the SOC creates a flux that pierces a plaquette in synthetic space which cannot be gauged away~\cite{Celi-2014}. 
This has been explored in quasi-one-dimensional ladder models (with $2S+1$ legs) through theoretical studies of interacting bosons~\cite{Orignac-2001,Dhar-2012,Dhar-2013,Petrescu-2013,Piraud-2015} or fermions~\cite{Po-2014,Zeng-2015,Barbarino-2015,Ghosh-2016} in the presence of a flux.

In Ref.~\onlinecite{Pixley-2016} the model in Eq.~(\ref{eqn:bh}) was studied directly in two real spatial dimensions using a Gutzwiller variational wave function~\cite{Rokshar-1991} on a square lattice, with SOC only along one of the two directions. Due to the presence of SOC, the superfluid phase at weak coupling is a striped superfluid~\cite{Zhai2010, Wu2010, Zhang2011, Li2012} with spin density wave order, which is suppressed for increasing $U_0/t$. At moderate $U_0/t$ the spin density wave order is destroyed and the superfluid condenses at a non-zero momentum only. Owing to strong lattice effects, this occurs on the edge of the Brilliouin zone~\cite{Pixley-2016,Hurst-2016}. Finally, a Mott transition occurs at even larger $U_0/t$. If we consider these results in the present context by considering the one dimensional model in Eq.~(\ref{eqn:bh}), the superfluid wavefunction is qualitatively captured by the Gutzwiller or Gross-Pitaevskii approximations~\cite{Hurst-2016}. However, because of strong quantum fluctuation effects in one dimension, we expect that the striped superfluid would be better described by a three component Luttinger liquid (similar to the pseudospin-1/2 case, Ref.~\onlinecite{Po-2014}).
Nonetheless, for sufficiently large interactions, the charge degrees of freedom in the Luttinger liquid will become gapped out leading to a bosonic Mott insulating phase. In this strong coupling limit, the appropriate degrees of freedom are local moments, governed by an effective Hamiltonian describing the limit of large $U_0/t$ only, that we now proceed to derive.

\begin{figure}[t]
\centering
\begin{minipage}{.5\textwidth}
  \centering
  \includegraphics[width=0.95\linewidth]{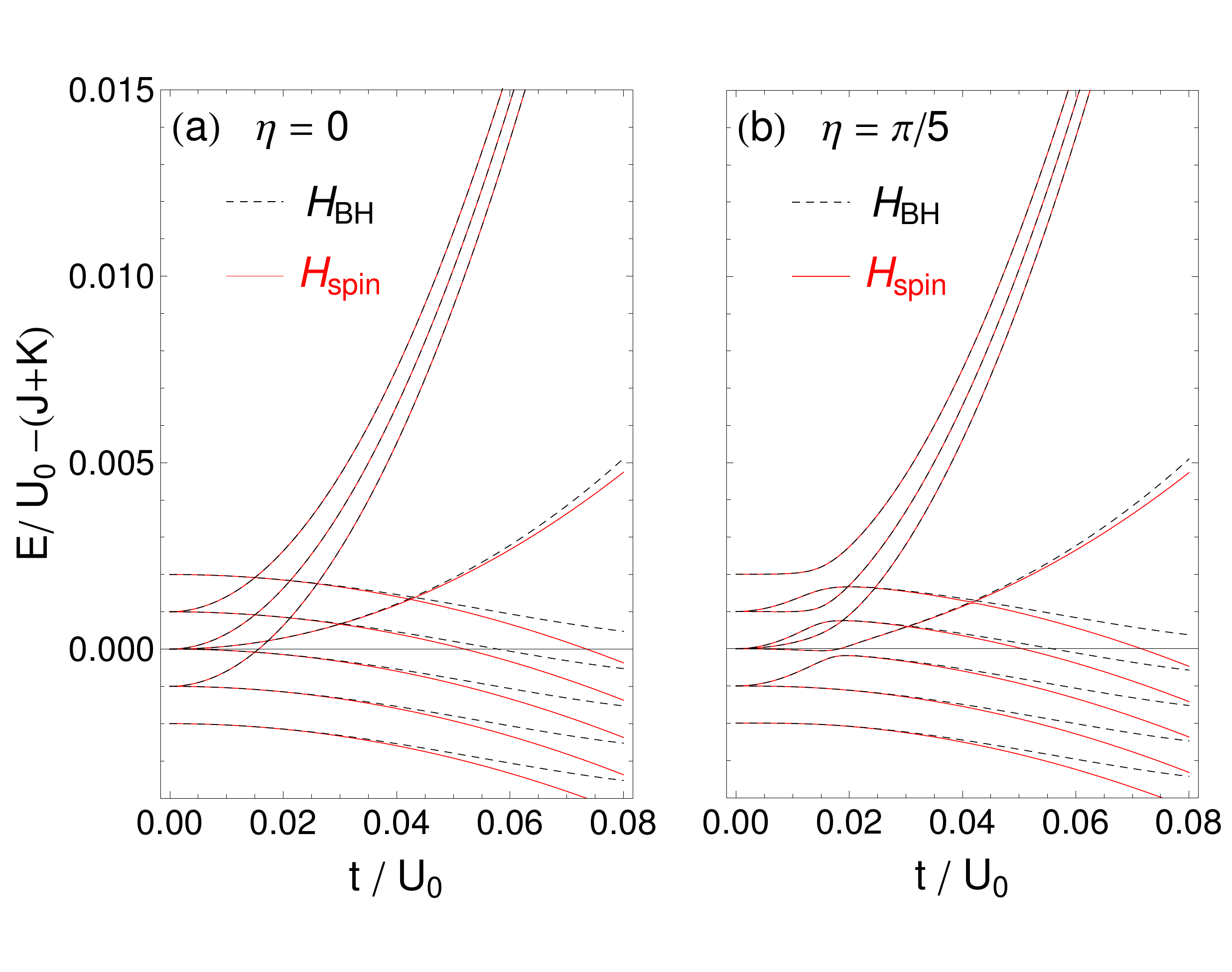}
\end{minipage}
\caption{Comparison of the low-energy eigenvalues of $H_{\rm BH}$ at unit filling with the spectrum of $H_{\rm spin}$ for two lattice sites, shown as a function of $t/U_0$ for (a) uniform field, and (b) SOC with pitch angle $\eta = \pi/5$. In both cases we take $h = 0.001 U_0$ and $U_2 = -0.1 U_0$. The spectra are indistinguishable for sufficiently small $t/U_0$, and the SOC causes avoided level crossings in the spectrum in the regime $t^2/U_0 \sim h$. At larger $t$, as the spectra begin to differ quantitatively, the level ordering nonetheless remains identical.} 
\label{fig:ed_compare}
\end{figure}

In the Mott limit of commensurate (here, odd integer) filling and $U_0,U_2 \gg t$, and in the absence of SOC ($h=0$),  Eq.~(\ref{eqn:bh_rotate}) is effectively described (perturbatively to order $t^2/U$) by the bilinear-biquadratic chain~\cite{Imambekov-2003} (ignoring the constant energy shift)
\begin{equation}
H_{\mathrm{spin}}(h = 0) = \sum_i \tilde{J} {\bf S}_i\cdot{\bf S}_{i+1}+ \tilde{K}({\bf S}_i\cdot{\bf S}_{i+1})^2,
\end{equation}
where now ${\bf S}_i$ denotes a spin-1 operator, $\tilde{J}$ and $\tilde{K}$ depend on the filling $N=2n+1$ of the Mott lobe and are given by~\cite{Imambekov-2003}
\begin{eqnarray}
-\frac{\tilde{J}}{t^2}&=&\frac{2(15+20n+8n^2)}{15(U_0+U_2)}-\frac{16(5+2n)n}{75(U_0+4U_2)},
\label{eqn:J}
\\
-\frac{\tilde{K}}{t^2}&=&\frac{2(15+20n+8n^2)}{45(U_0+U_2)}+\frac{4(1+n)(3+2n)}{9(U_0-2U_2)} \nonumber \\
& & + \frac{4n(5+2n)}{225(U_0 + 4 U_2)}.
\label{eqn:K}
\end{eqnarray}
Following standard notation, we parametrize $\tilde{J}$ and $\tilde{K}$ on the circle 
\begin{eqnarray}
\tilde{J} &=& J \cos \theta,
\label{eqn:J2}
\\
\tilde{K} &=& J \sin \theta,
\label{eqn:K2}
\end{eqnarray}
 and $J$ is the unit of energy.
The antiferromagnetic chain ($\theta=0$) is gapped (whereas the corresponding spin-1/2 model is gapless), and characterized by a non-zero hidden, string-like, order parameter.
The ground state is very efficiently described by a matrix product state wavefunction, and exhibits symmetry-protected topological order~\cite{AKLT-1987,White-1993}. The ferromagnetic case ($\theta=\pi$), on the other hand, is gapless, ordered, and topologically trivial. Tuning the biquadratic interaction enriches the problem even further: for sufficiently negative interaction ($5 \pi/4 < \theta < 7 \pi/4$) the translational symmetry of the model is spontaneously broken, and the ground state is dimerized~\cite{Klumper-1989,Sorensen-1990,Matteo-2005}. In contrast, for a large \emph{positive} biquadratic interaction ($\pi/4 < \theta <  \pi/2$) there is a spin-quadrupolar phase with gapless excitation modes at momenta $q=0,\pm 2\pi/3$  (see Ref.~\onlinecite{Lauchli-2006} and references therein). The quantum phase transitions separating these phases are quite interesting, and have been described by various conformal field theories~\cite{Affleck-1987} and Bethe ansatz solutions~\cite{Uimin-1970,Lai-1974,Sutherland-1975,Takhtajan-1982,Babujian-1982}.

We determine the effective spin Hamiltonian for $h\neq 0$ in the odd integer Mott lobes using a Schrieffer-Wolf transformation~\cite{Schrieffer-1966} and then a projection. The final Hamiltonian is $H_{\mathrm{spin}}=P_s H' P_s$, where $P_s$ projects into the subspace of filling $N$ (e.g. focusing on the first Mott lobe $P_s$ projects into the singly occupied subspace) and
\begin{eqnarray}
H' = e^{-O } H e^{ O} = H - [O,H] + \frac{1}{2!} [O,[O,H]] + \cdots
\end{eqnarray}
where $O$ is chosen such that $H_t - [O,H_U] = 0$. Using properties of projectors and the fact that $H_{h}$ does not change the particle number subspace, i.e. $P_s H_h P_d=P_dH_hP_s=0$ (where $P_d$ projects into the $N+1$ occupied subspace), we find $O=(P_sH_tH_U^{-1}P_d-P_dH_U^{-1}H_tP_s)$. Thus, the form of $O$ is unaffected by the presence of $H_h$, which implies that the form of $\tilde{J}$ and $\tilde{K}$ are unchanged from Eqs. (\ref{eqn:J}) and (\ref{eqn:K}). As a result we find
$P_s [O,H_h] P_s=0$ while 
\begin{eqnarray}
P_s[O,[O,H_h]]P_s\sim t^2 h/U^2   \ll t^2/U
\label{eqn:correction}
\end{eqnarray} 
where $U = \mathrm{min}(U_0,|U_2|)$.
Therefore, to leading order in $t^2/U$ we obtain  
\begin{equation}
H_{\mathrm{spin}}= \sum_i \tilde{J}{\bf S}_i\cdot{\bf S}_{i+1}+ \tilde{K}({\bf S}_i\cdot{\bf S}_{i+1})^2+{\bf h}_i\cdot {\bf S}_i,
\label{eqn:spin_model}
\end{equation}
where  $\tilde{J}(<0)$ and  $\tilde{K}$ are parametrized on the circle as in defined in Eqs.~(\ref{eqn:J2}) and ~(\ref{eqn:K2}) respectively.
We stress that even for $h/J$ on the order of one the Hamiltonian in Eq.~(\ref{eqn:spin_model}) is still valid as the next order correction in Eq.~(\ref{eqn:correction}) goes like $\sim(h/J) t^4 / U^3$. The Hamiltonian in Eq.~(\ref{eqn:spin_model}) has a $Z_2$ symmetry about the $x-y$ plane and is invariant under the reflection $S_i^z\rightarrow -S_i^z$.
For typical spin-1 bosonic atoms $|U_2| \simeq 0.05 |U_0|$ (see Ref.~\onlinecite{2013_SKurn_Ueda_RMP}).

In the frame spatially co-rotating with the helical magnetic field, the SOC induces
a Dzyaloshinskii-Moriya (DM) type interaction~\cite{Radic2012,mott_2d_cole}. To see this, we choose a generator $G=-i\eta \sum_r r S_r^z$ and apply the canonical transformation to the spin model $H'_{\mathrm{spin}}=e^GH_{\mathrm{spin}}e^{-G}$. This ``unwinds'' the spiral field $e^G {\bf h}_i\cdot{\bf S}_ie^{-G}=hS_i^x$ and induces a cross product  in the product of spin operators
\begin{eqnarray}
e^G ({\bf S}_i \cdot {\bf S}_{i+1})e^{-G}=\cos(\eta)(S_i^xS_{i+1}^x+S_i^yS_{i+1}^y) 
\nonumber
\\
+ \sin(\eta)({\bf S}_i \times {\bf S}_{i+1})\cdot \hat{z}+S_i^zS_{i+1}^z.
\label{eqn:dmint}
\end{eqnarray}
This expression shows that in the rotating frame the SOC breaks the $xy-z$ symmetry and induces a DM interaction.

Finally, we further validate this effective low energy Hamiltonian by comparing exact diagonalization of the Bose-Hubbard model in Eq.~(\ref{eqn:bh_rotate}) at unit filling and the effective spin model we have derived in Eq.~(\ref{eqn:spin_model}) for two lattice sites. For the latter, we can also make exact statements about the energy eigenstates. For $h=0$, these are also clearly eigenstates of total $S_t = S_1 + S_2$, and ${\bf S}_1 \cdot {\bf S}_2 = \frac{1}{2} S_t (S_t + 1) - 2$. So, of the possible multiplets $S_t = 0,1,2$, $S_t = 2$ maximizes ${\bf S}_1 \cdot {\bf S}_2$, reflecting the ferromagnetic tendency of this term. On the other hand, for $S_t = 0$, ${\bf S}_1 \cdot {\bf S}_2 = -2$, so $({\bf S}_1 \cdot {\bf S}_2)^2 = 4$. For a dominant, negative $\tilde{K}$, this total spin singlet has the lowest energy, reflecting the tendency there for dimerization. A nonzero uniform field, $h \neq 0$, $\eta = 0$ yields the term $h\sum_i S_i^x$ that commutes with the total spin, and the multiplets are simply Zeeman split. As soon as the field is nonuniform ($\eta \neq 0$), this term no longer commutes: the Hamiltonian is frustrated.
Fig.~\ref{fig:ed_compare} shows
 a numerical comparison of the two-site spectra for the two models. For the full spin-1 Bose-Hubbard model, the states shown have weight almost exclusively in the single-occupancy subspace, 
up to corrections of order 1 \%
because of the large gap ($U_0 + U_2$) to double occupancy. The arrangement of states is sensible in terms of the $S_t$ eigenstates described. Here, as a representative case we take  $U_2 = -0.1 U_0$, resulting in $\tilde{K}/\tilde{J} = 5/6$, favoring first $S_t = 2$ then $S_t = 0$. For $\eta \neq 0$, we now see the effect of frustration in the appearance of several avoided crossings in the region $h \sim \tilde{J},\tilde{K}$.
This result provides strong evidence that the perturbative treatment to derive the effective Hamiltonian in the presence of a SOC is sufficient and higher order terms [as in Eq.~(\ref{eqn:correction})] are negligible. Next, however, to understand the behavior of this model for more than two sites, we must apply more sophisticated techniques.

 \begin{figure}[t]
\centering
\begin{minipage}{.25\textwidth}
  \centering
  \includegraphics[width=0.7\linewidth,angle=-90]{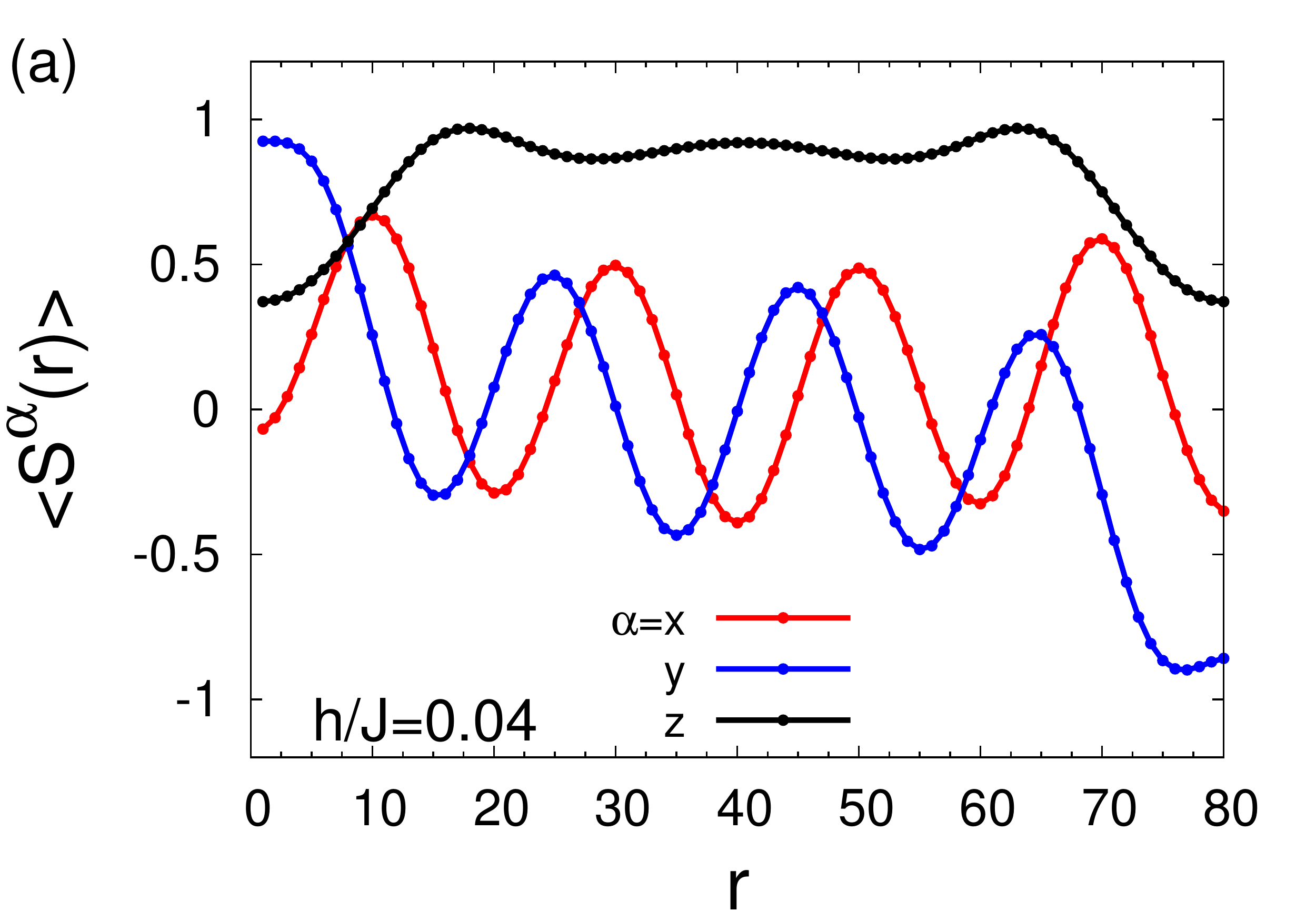}
\end{minipage}%
\begin{minipage}{.25\textwidth}
  \centering
  \includegraphics[width=0.7\linewidth,angle=-90]{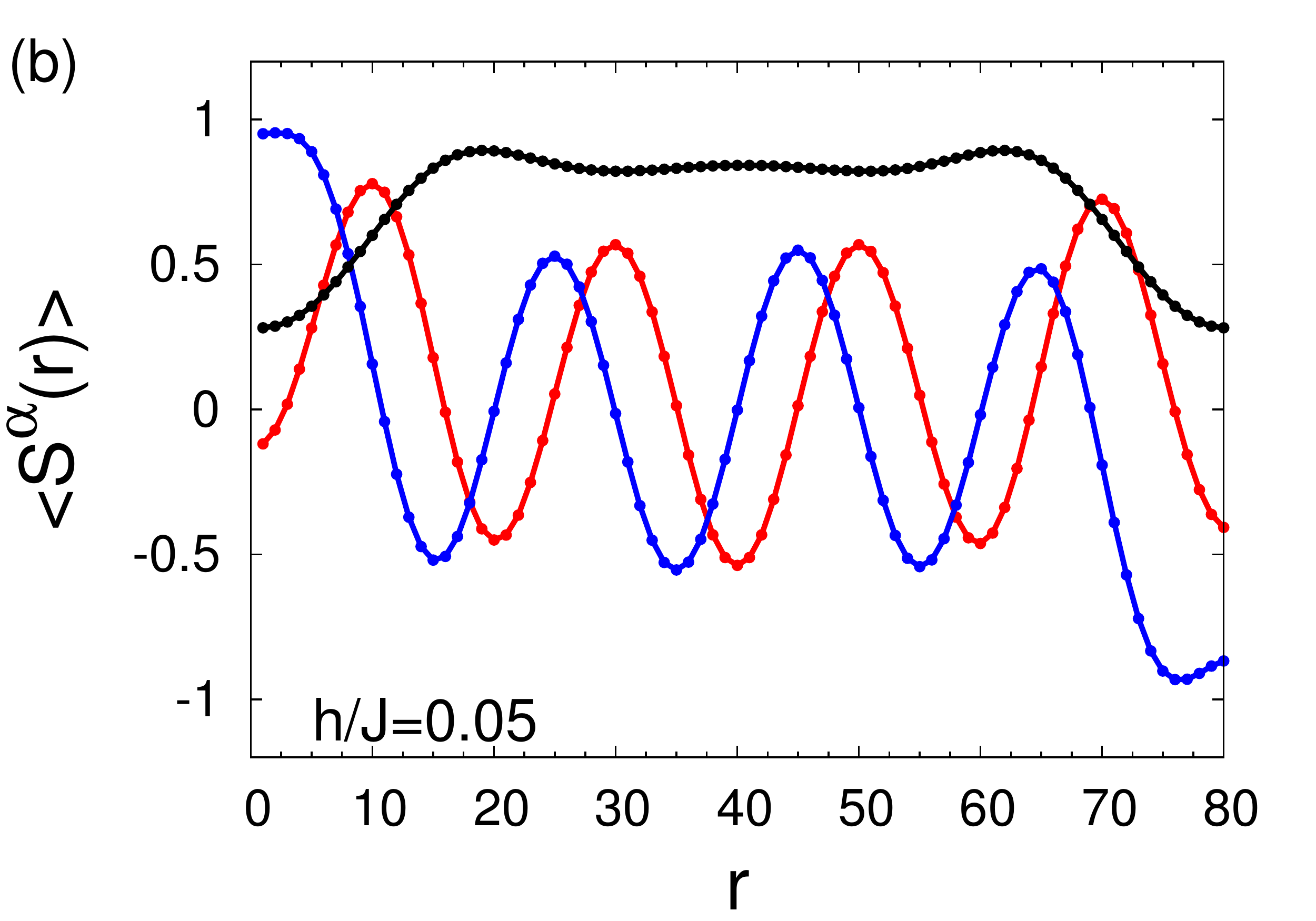}
\end{minipage}
\newline
\centering
\begin{minipage}{.25\textwidth}
  \centering
  \includegraphics[width=0.7\linewidth,angle=-90]{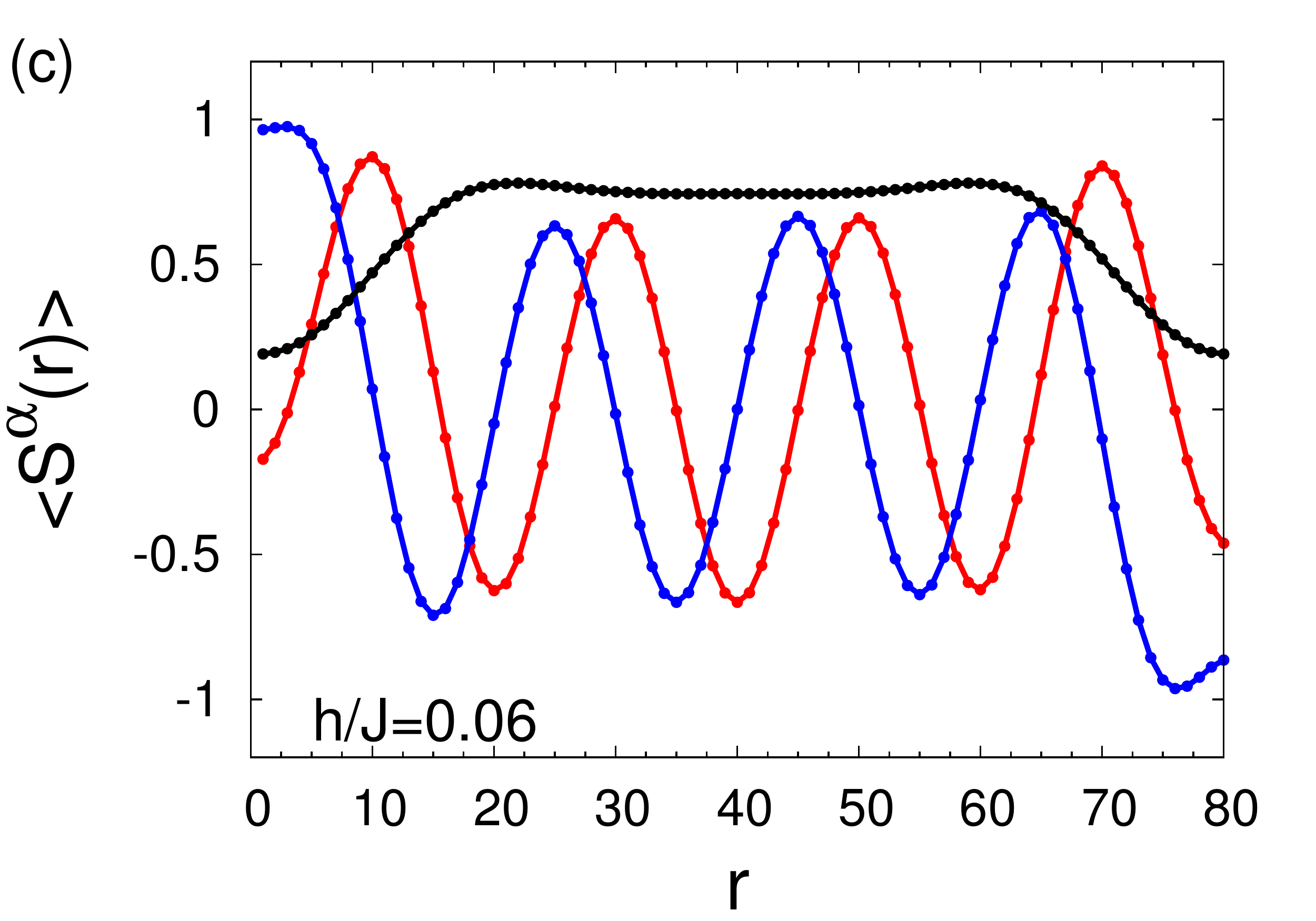}
\end{minipage}%
\begin{minipage}{.25\textwidth}
  \centering
  \includegraphics[width=0.7\linewidth,angle=-90]{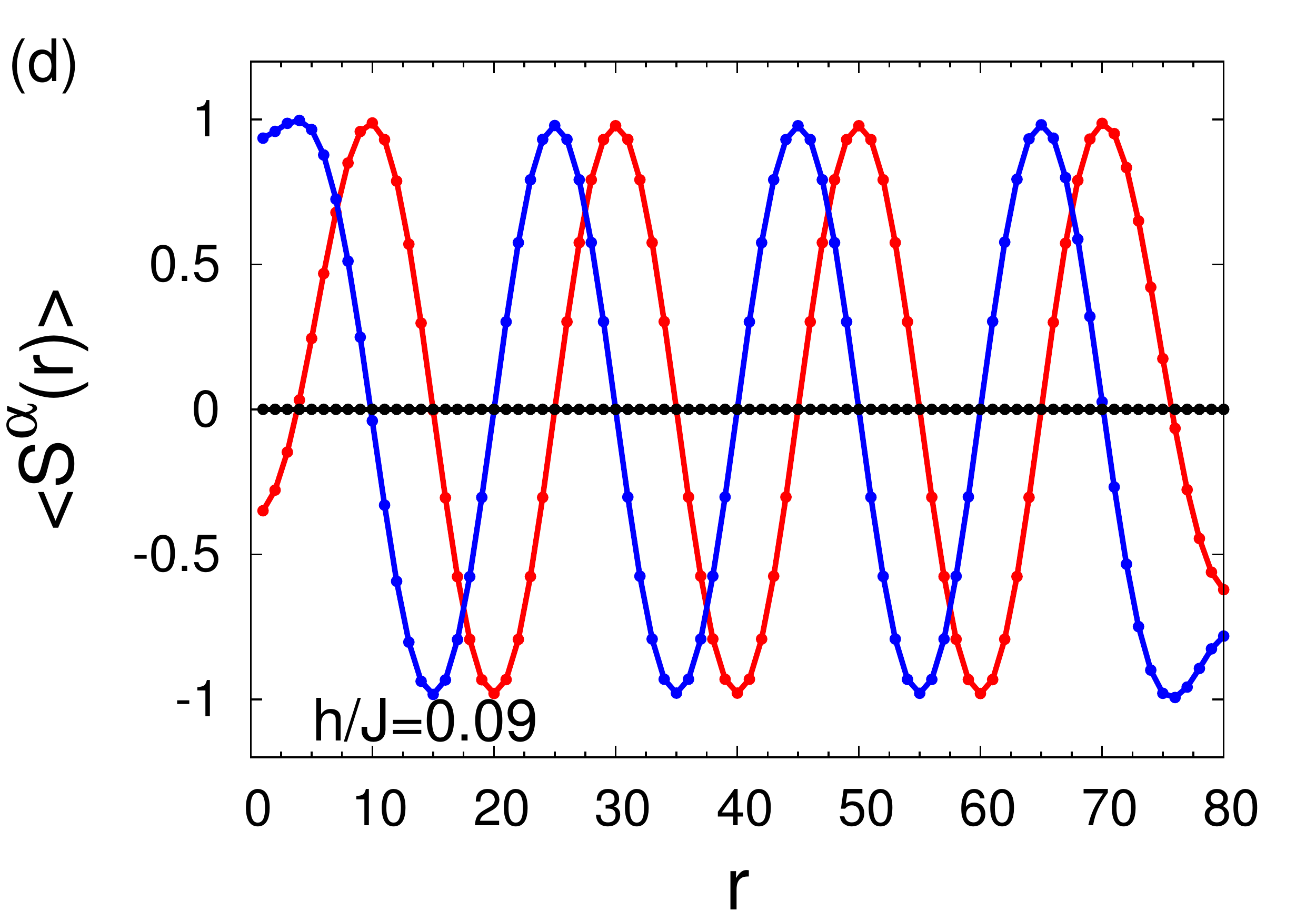}
\end{minipage}
\caption{Average spin along the chain in the SFM (a), (b), and (c), as well as the SPM (d) phase computed using DMRG with chain length $L=80$, bond dimension $M=50$, and $\theta=1.1\pi$. For a pitch $\eta=\pi/10$ the spiral is commensurate with $L$ and there are large finite size effects for small $h$ near the chain boundaries, we have also confirmed this behavior in the classical model [see Figs.~\ref{fig:classical_panel} (a) and (b)]. Away from the chain boundaries we find $\langle S^x(r)\rangle \propto \cos(\eta r)$ and $\langle S^y(r)\rangle \propto \sin(\eta r)$ consistent with the Landau theory (see Sec.~\ref{subsec:LT}). For moderate fields the boundary effects are weak and $\langle S^z(r)\rangle$ is roughly constant in the center of the chain. Crossing the critical field $h_c(\theta=1.1\pi)/J = 0.0798 (3)$ we find $M_z=0$.} 
\label{fig:mx}
\end{figure}

In the remainder of the paper, we study the effective spin Hamiltonian $H_{\mathrm{spin}}$ (valid in the strong coupling limit).
Even though the spin Hamiltonian in Eq.~(\ref{eqn:spin_model}) is well defined for $\tilde{J}>0$ and $\tilde{J}<0$, we focus on the physics relevant to strongly coupled bosons and limit our attention to a ferromagnetic nearest neighbor coupling $\tilde{J}<0$, (i.e., restricting ourselves to the regime $\pi/2 \le \theta \le 3\pi/2$). 
We propose to access the phases contained in the spin Hamiltonian in Eq.~(\ref{eqn:spin_model}) in experiment dynamically as it is straightforward to first enter the Mott phase with a spin polarized (i.e. ferromagnetic) gas and then to adiabatically apply the SOC fields, retaining the low entropy of the initial spin polarized gas.
We use DMRG to compute the phase diagram as a function of $h$ and $\theta$, for a fixed pitch $\eta=\pi/10$ (unless otherwise stated). 
In the absence of any good quantum number to be exploited, the simulations become rapidly quite expensive and slow. 
We stress that the $Z_2$ symmetry related to the reflection about the $x-y$ plane is a {\it{anti-}}unitary operation (indeed it changes the sign to a single spin component, $S_z$). Therefore it doesn't fall within the local point-wise unitary symmetries whose quantum numbers can be easily encoded in the tensor network structure. The main gain one could get out of it is to exploit the reality of the Hamiltonian (and thus the existence of a basis of real-valued eigenvectors) to slightly reduce the computational costs. (This is, however, a minor gain compared to the ones usually achieved by conserved quantum numbers).
Therefore we limited our investigations to bond dimensions of a few hundreds: we explicitly checked that the discarded probability in the renormalization process was low enough, and regions where this was not the case were left open for further investigations.
We supplement the numerically exact DMRG calculations with 
a description of the spiral ferromagnetic phase by solving the model in the classical approximation and constructing a phenomenological Landau theory.

\section{Spiral Ferromagnet}
\label{sec:sf}
 \subsection{DMRG}

In the absence of the field ($h=0$), and $0.5\pi \le \theta \lesssim 1.25\pi$, the model has a largely degenerate ground manifold (namely, all $2L+1$ states with maximal total spin $S^{\mathrm{TOT}} = L$): in particular, the DMRG simulations will tend to select out a minimally entangled state from such a manifold, in this case it is a product state of parallel spins (e.g., ``all spins up'', $S^z_j = +1 \  \forall j$, but not necessarily). 
Turning on a weak field, the ground degeneracy is reduced to a two-fold one (related to the $\mathbb{Z}_2$ reflection symmetry $S_z \rightarrow -S_z$ mentioned before) and a spontaneous symmetry breaking could still take place. The system indeed acquires a spiral configuration in $\langle S^x(r) \rangle$ and $\langle S^y(r) \rangle$, while reducing the magnitude of the parallel-oriented $\langle S^z(r) \rangle$ and the ground state becomes a spiral ferromagnet (SFM) (see Fig.~\ref{fig:mx}).
At very weak fields we find a strong finite size effect near the chain boundaries, but for moderate field strengths this finite size effect is suppressed and $\langle S^z(r) \rangle$ becomes essentially constant in the center of the chain, as shown in Figs.~\ref{fig:mx} (a), (b), and (c).  Upon increasing $h$ further, the system finds it energetically favorable to orient all the spins within the $xy$-plane (i.e., $\langle S^z(r) \rangle$ goes to zero) and let them follow the rotating magnetic field: the ground state is then a spiral paramagnet (SPM). 
Thus $M_z \simeq \langle S^z(L/2) \rangle$ could apparently serve as an order parameter for the transition. 

\begin{figure}[t]
\centering
\begin{minipage}{.5\textwidth}
  \centering
  \includegraphics[width=0.7\linewidth,angle=-90]{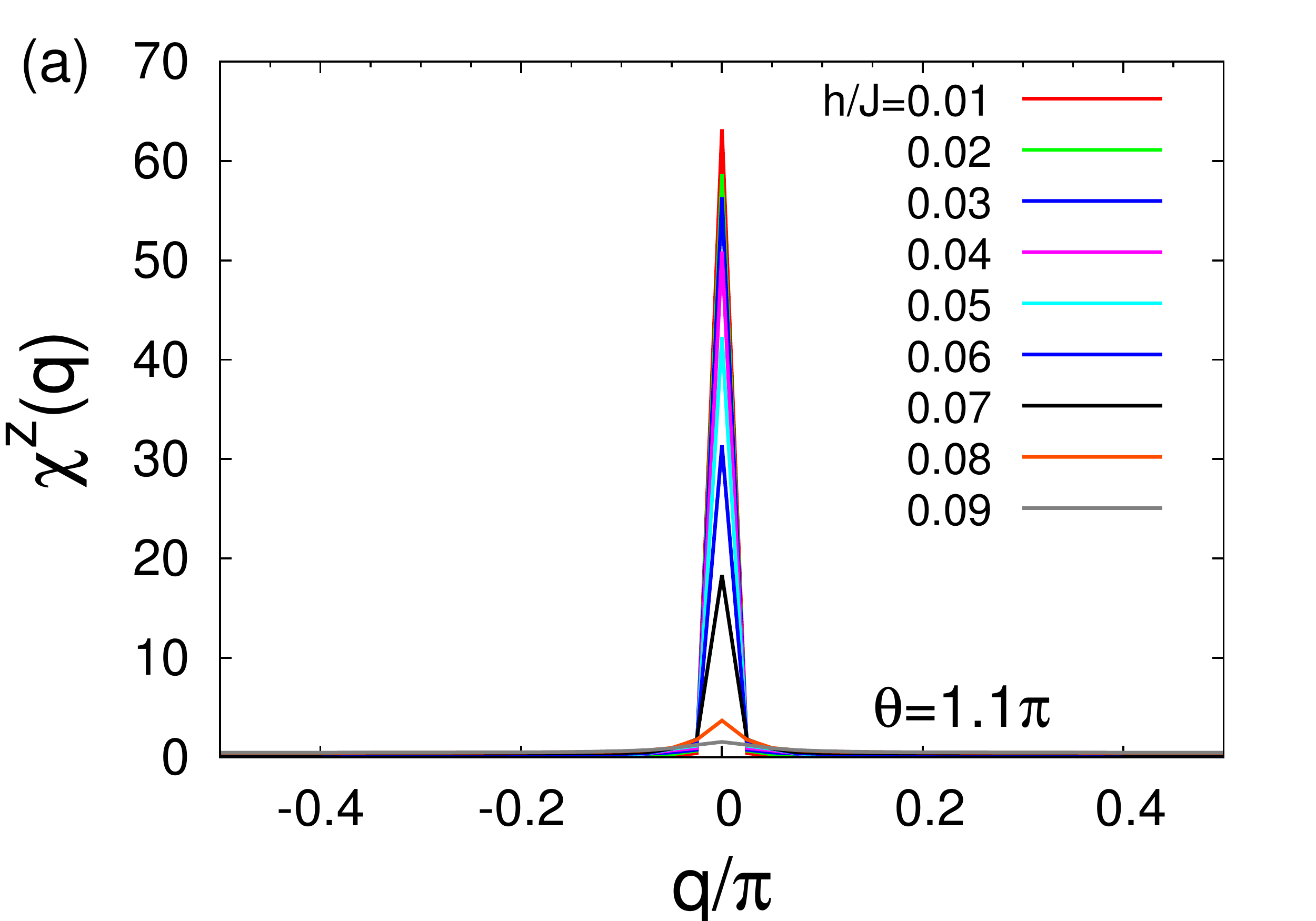}
\end{minipage}%
\newline
\begin{minipage}{.5\textwidth}
  \centering
  \includegraphics[width=0.7\linewidth,angle=-90]{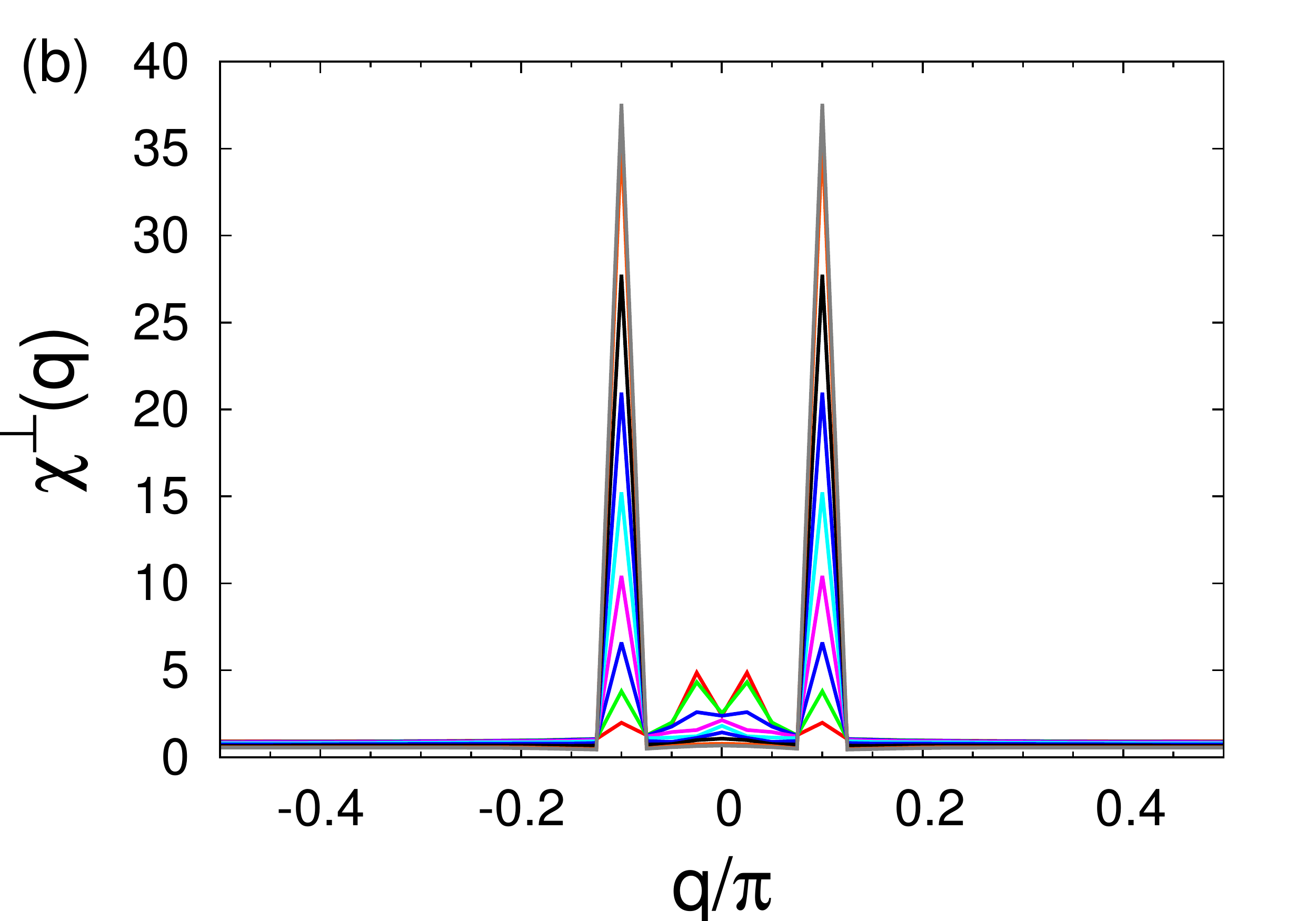}
\end{minipage}
\caption{Spin susceptibility computed from DMRG as a function of momentum $q$ in the longitudinal (a) and transverse (b) directions with $L=80$, $M=50$, $\theta=1.1\pi$, and the legend is shared across both figures. In the SFM phase $\chi^z(q)$ is strongly peaked at $q=0$ with a peak height monotonically decreasing with increasing $h$ going to zero at $h_c$. The non-zero SOC produces peaks in $\chi^{\perp}(q)$ at $q=\pm \eta(=\pm 0.1\pi)$ and the peak continuously increases as a function of $h$. } 
\label{fig:Sz_q_h}
\end{figure}

We stress here that the two-fold degeneracy of the SFM
at finite $h$ would be exact only in the thermodynamic
limit, while finite-size effects will slightly split it between
Òcat-state" superpositions of definite $\mathbb{Z}_2$ symmetry (just
like in the transverse-field Ising model).
This would therefore lead to a null $\langle S^z (L/2) \rangle$, thus preventing the correct identification of the transition point.
While the well-known DMRG bias towards minimally entangled states
(due to inevitably finite bond dimensions) would tend to numerically induce the symmetry breaking even in a finite system, this only happens when the degeneracy splitting is small enough.
Since such a splitting is vanishing with system size slower and slower the closer we are to the SFM-SPM transition (again, like in the Ising model), the use of the local order parameter to pinpoint the quantum critical point would be quite unreliable.
In order to avoid such problems, we study the spin correlation functions $C^z(r,r^\prime)=\langle S^z(r)S^z(r^\prime) \rangle$ and $C^{\perp}(r,r^\prime) = \langle S^x(r)S^x(r^\prime) \rangle + \langle S^y(r)S^y(r^\prime) \rangle$.
To understand the ordering pattern of spins we consider the Fourier transforms
 \begin{equation}
 \chi^{\alpha}(q) = \frac{1}{L}\sum_{r, r'} e^{i q(r-r')}C^{\alpha}(r,r'),
 \end{equation}
see Fig.~\ref{fig:Sz_q_h}. 
In the SFM phase we find $\chi^z(q)$ is peaked at $q=0$ and $\chi^{\perp}(q)$ is peaked at $q=\pm \eta (=\pm 0.1\pi)$: we find the peak height of $\chi^z(q)$ $[\chi^{\perp}(q)]$ decreases [increases] with increasing $h$.
Thus the appropriate magnetic order parameter that we can correctly extract from the DMRG data is given by $M_z = \lim_{L \rightarrow \infty} \sqrt{\chi^z(q=0)/L}$.

 \begin{figure*}[t]
\centering
\begin{minipage}{.5\textwidth}
  \centering
  \includegraphics[width=0.7\linewidth,angle=-90]{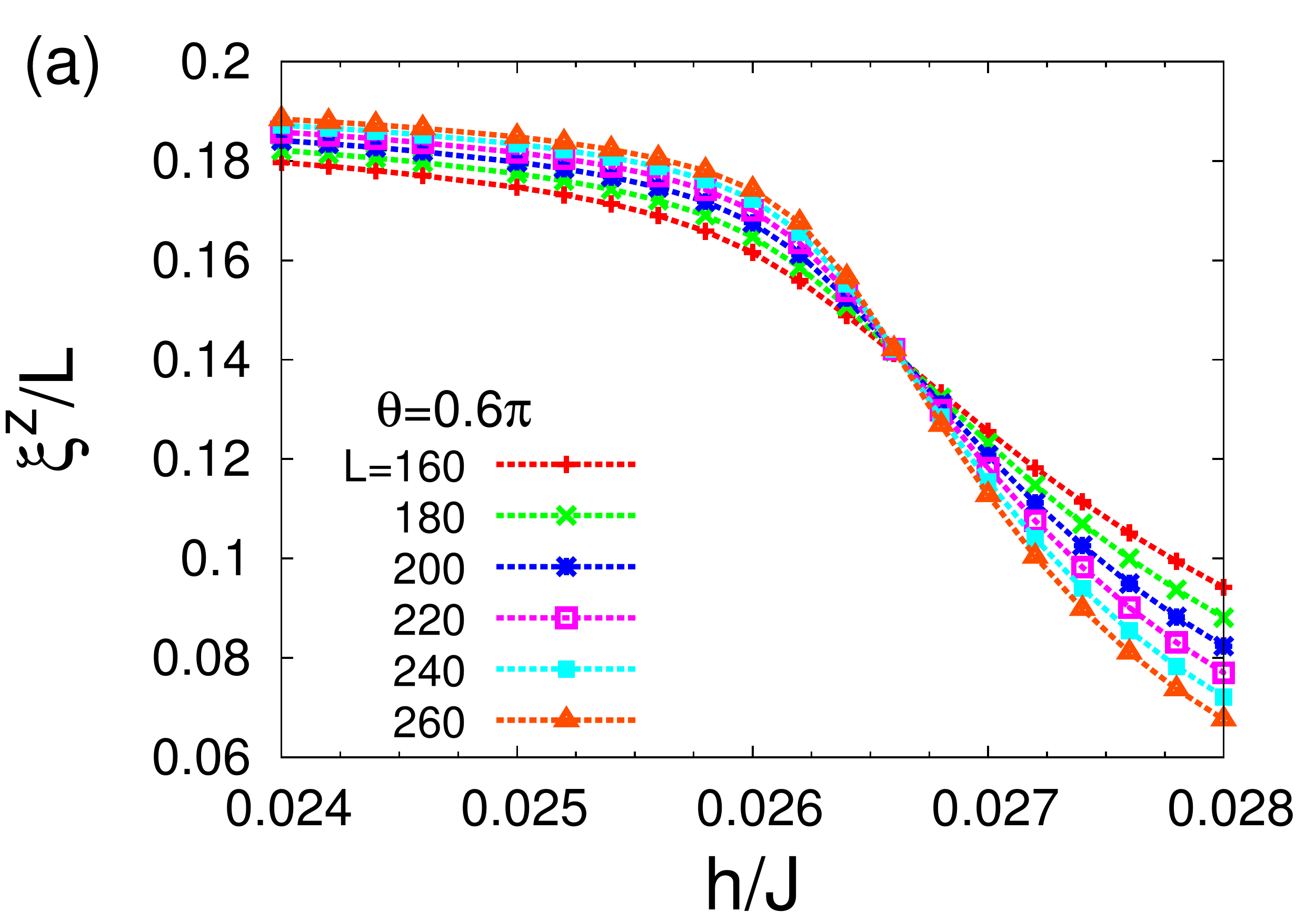}
\end{minipage}%
\begin{minipage}{.5\textwidth}
  \centering
  \includegraphics[width=0.7\linewidth,angle=-90]{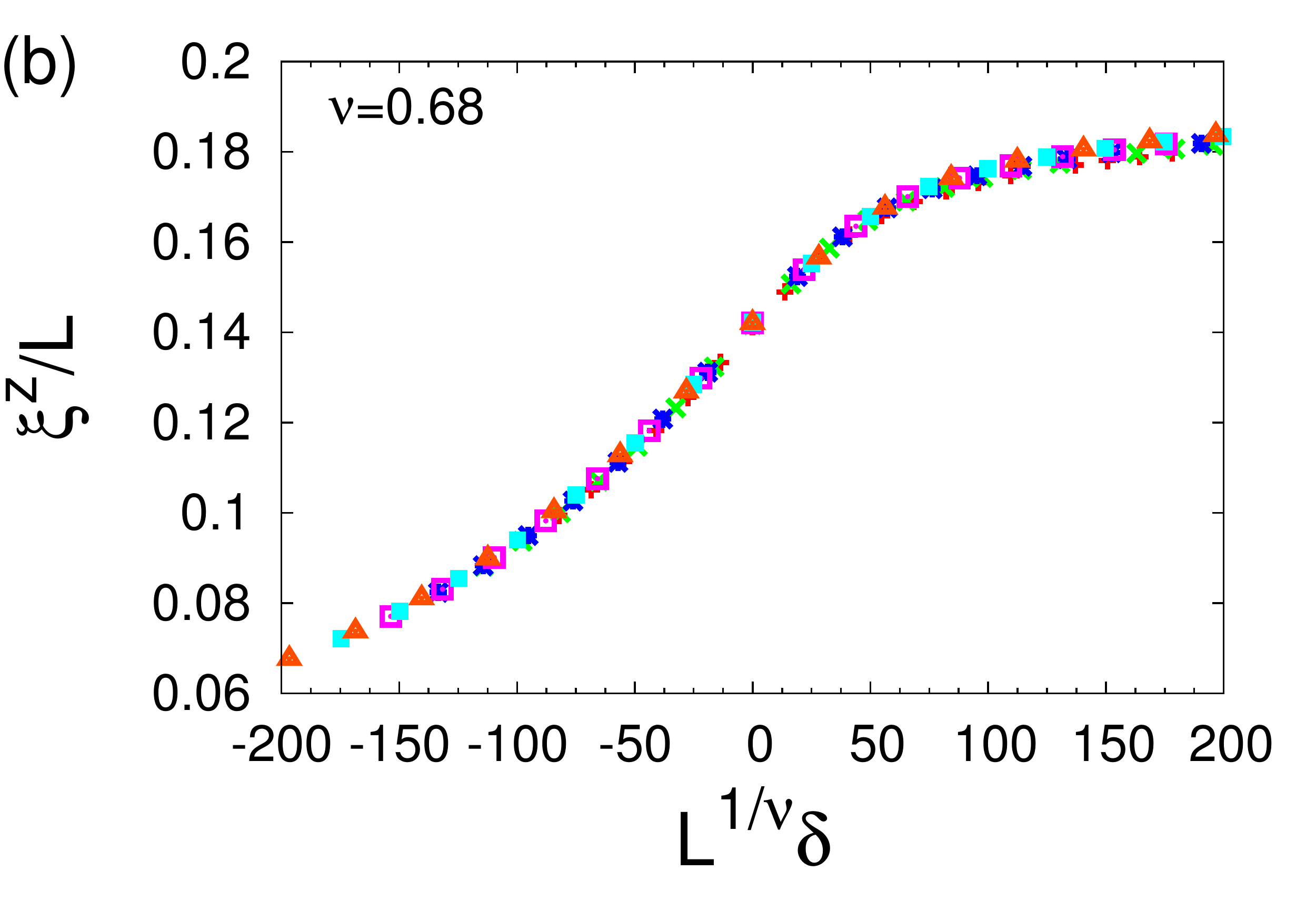}
\end{minipage}
\newline
\begin{minipage}{.5\textwidth}
  \centering
  \includegraphics[width=0.7\linewidth,angle=-90]{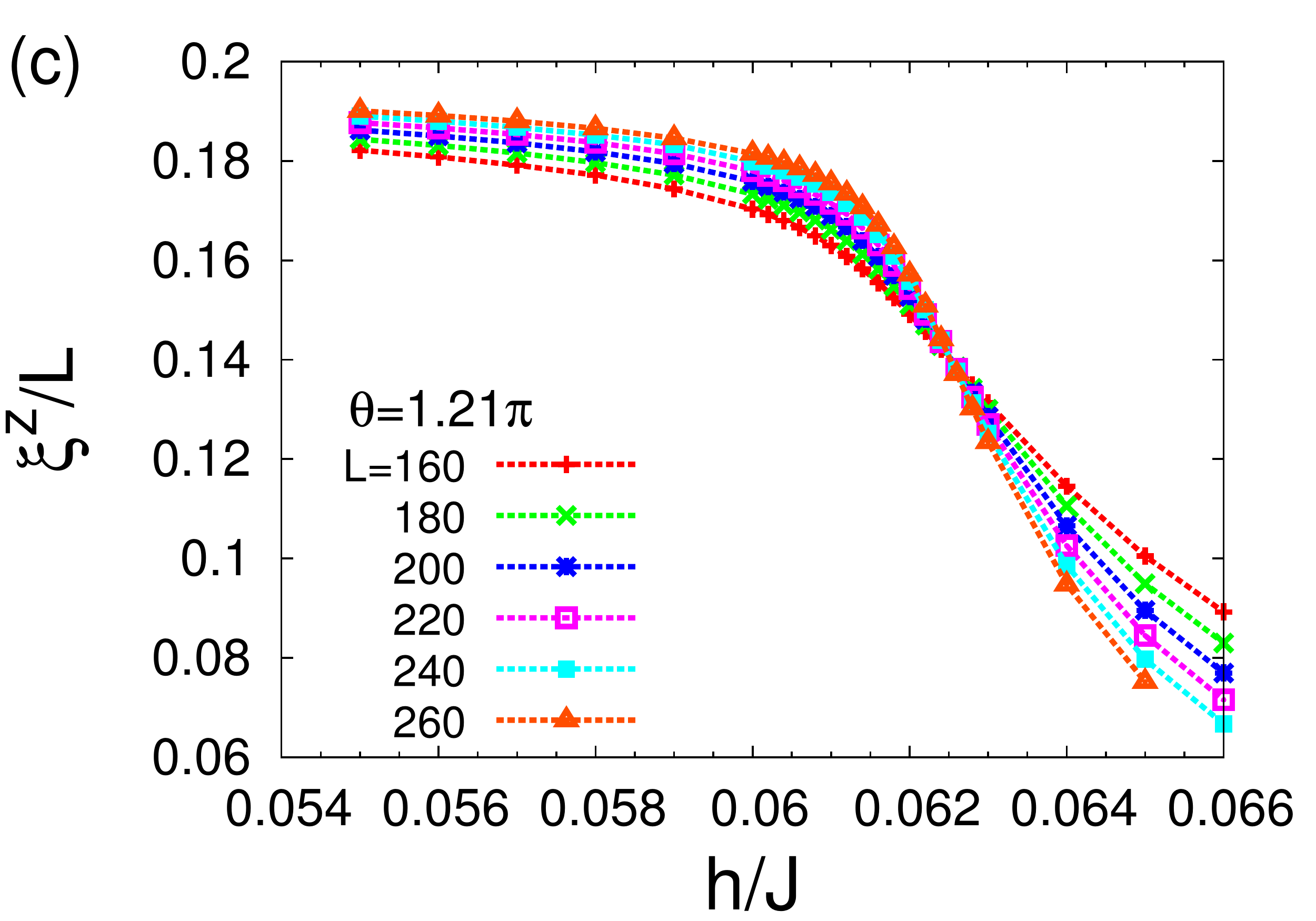}
\end{minipage}%
\begin{minipage}{.5\textwidth}
  \centering
  \includegraphics[width=0.7\linewidth,angle=-90]{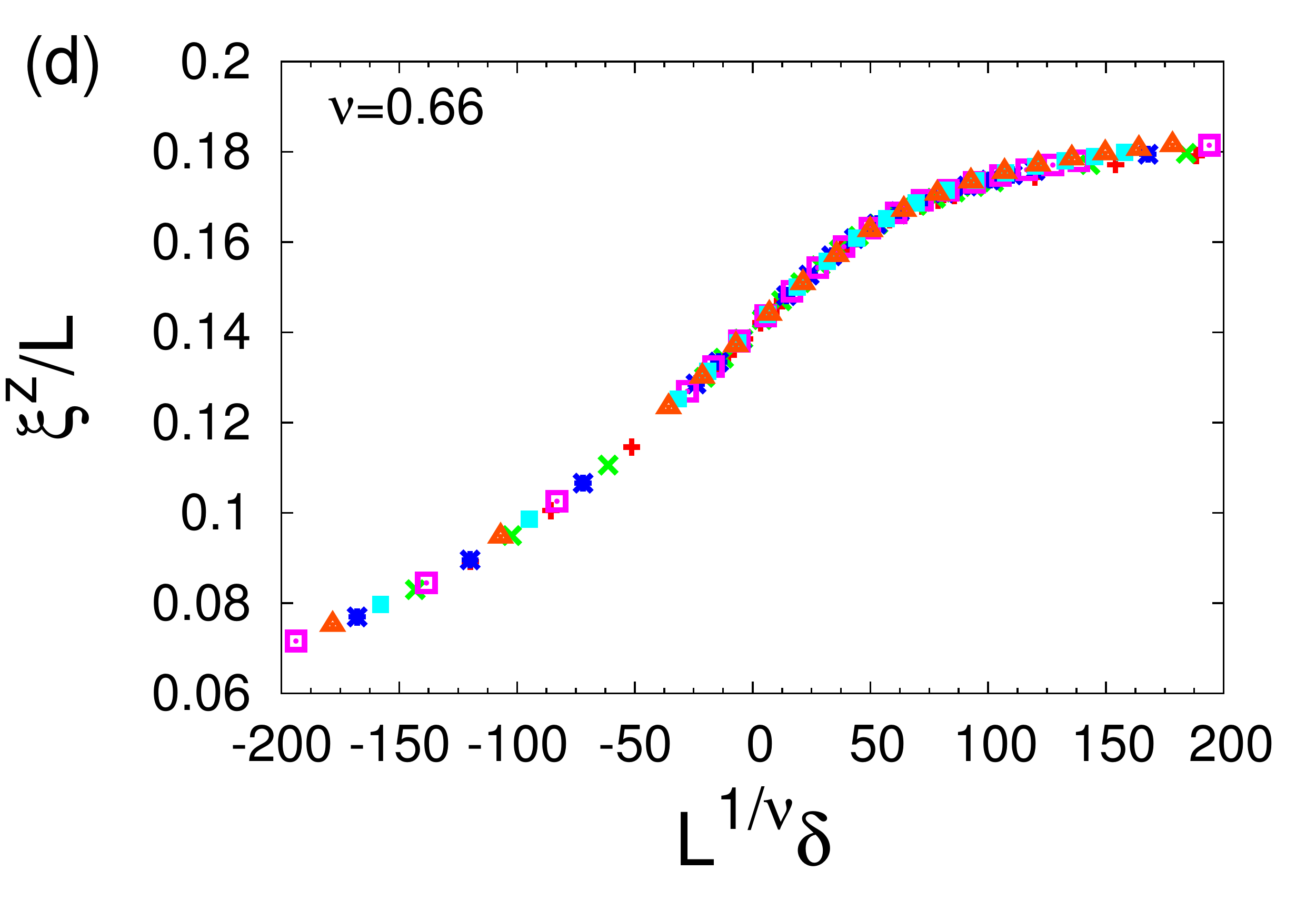}
\end{minipage}
\caption{Longitudinal correlation length $\xi^z$ defined in Eq. (\ref{eqn:xi}) for various system sizes as a function of $h$ for $\theta =0.6\pi$ (a) and (b) as well as $\theta=1.21\pi$ (c) and (d). We find a clear crossing in the data for $\xi^z/L$ versus $h$ for several $L$ and take this location as an unbiased estimate of $h_c$, we find $h_c(\theta=0.6 \pi)=0.0266(2)$ and $h_c(\theta=1.21\pi)=0.0625(3)$. In the vicinity of the transition we perform finite size scaling from Eq.~(\ref{eqn:scaling}) and collapse the data in terms of $L^{1/\nu}\delta$ and estimate the quality of collapse using a $\chi^2$ analysis~\cite{Kawashima-1993}. We find $\nu = 0.68 (8)$ for $\theta=0.6\pi$ and $\nu = 0.66 (6)$ for $\theta=1.21 \pi$.} 
\label{fig:xiL}
\end{figure*}

To determine the location of the quantum phase transition at $h_c$, separating the SFM and SPM phases we use finite size scaling with open boundary conditions as a function of $L$ and $h$. 
We define a correlation length from the second moment of the spin susceptibility~\cite{Campostrini-2014}
\begin{equation}
\xi^z = \sqrt{\frac{\sum_{r\neq L/2} (r-L/2)^2 C^z(r,L/2)}{2 \sum_{r\neq L/2} C^z(r,L/2)}}.
\label{eqn:xi}
\end{equation}
We only measure correlations from the center of the chain to minimize edge effects due to the open boundaries.
In the vicinity of the phase transition (in the thermodynamic limit) $\xi^z\sim |\delta|^{-\nu}$ where $\delta \equiv (h-h_c)/h_c$. However in our finite size numerics this divergence is cut off and rounded out, thus $\xi^z \sim L$ at $h=h_c$. Therefore, considering $\xi^z/L$ as a function of $h$ for various $L$ the data will cross at the critical field as shown in Figs.~\ref{fig:xiL} (a) and (c). As a result we can find the location of the critical point to high accuracy. Near the transition $\xi^z$ obeys single parameter scaling
\begin{equation}
\xi^z/L \sim  f(L^{1/\nu}\delta),
\label{eqn:scaling}
\end{equation}
where $f(x)$ is an arbitrary scaling function. As shown in Figs.~\ref{fig:xiL} (b) and (d) we find excellent data collapse of $\xi^z/L$ versus $L^{1/\nu}\delta$.
For $\theta=0.6\pi$ we find $\nu = 0.68 (3)$ and $\theta=1.21\pi$ yields $\nu = 0.66 (3)$. We have computed $\nu$ at several  other points along the SFM to SPM phase boundary (not shown) and find a critical exponent $\nu$ that ranges from $0.63(3)$ to at most $0.68(3)$, which all agree within the error bars. Based on the crossing and quality of collapse we conclude that the SFM-to-SPM transition is a second order continuous quantum phase transition, $\nu$ is independent of $\theta$ within numerically accuracy, and we find $\nu \approx 2/3$.

\begin{figure}[t]
\centering
\begin{minipage}{.5\textwidth}
  \centering
  \includegraphics[width=0.7\linewidth,angle=-90]{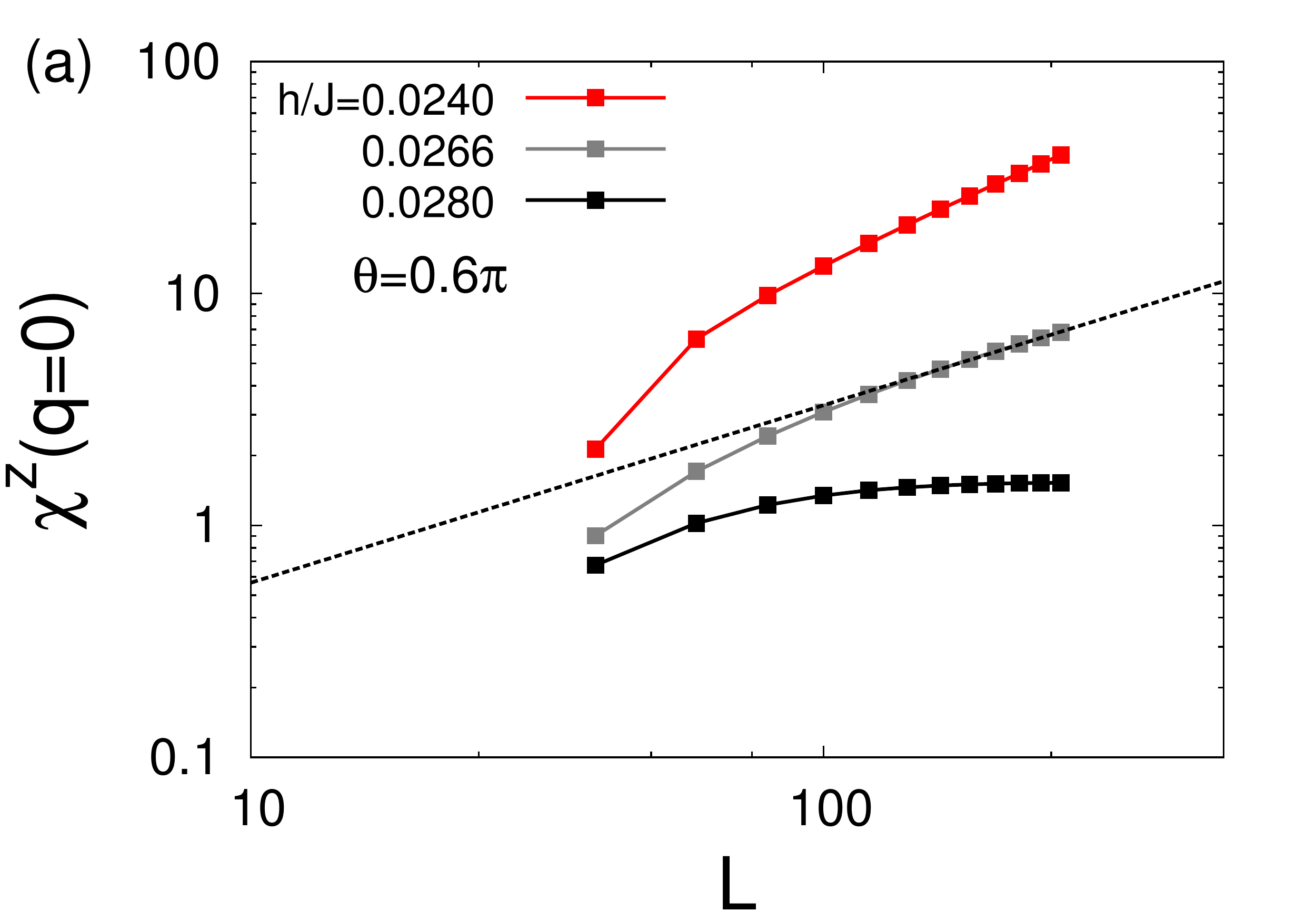}
\end{minipage}%
\newline
\begin{minipage}{.5\textwidth}
  \centering
  \includegraphics[width=0.7\linewidth,angle=-90]{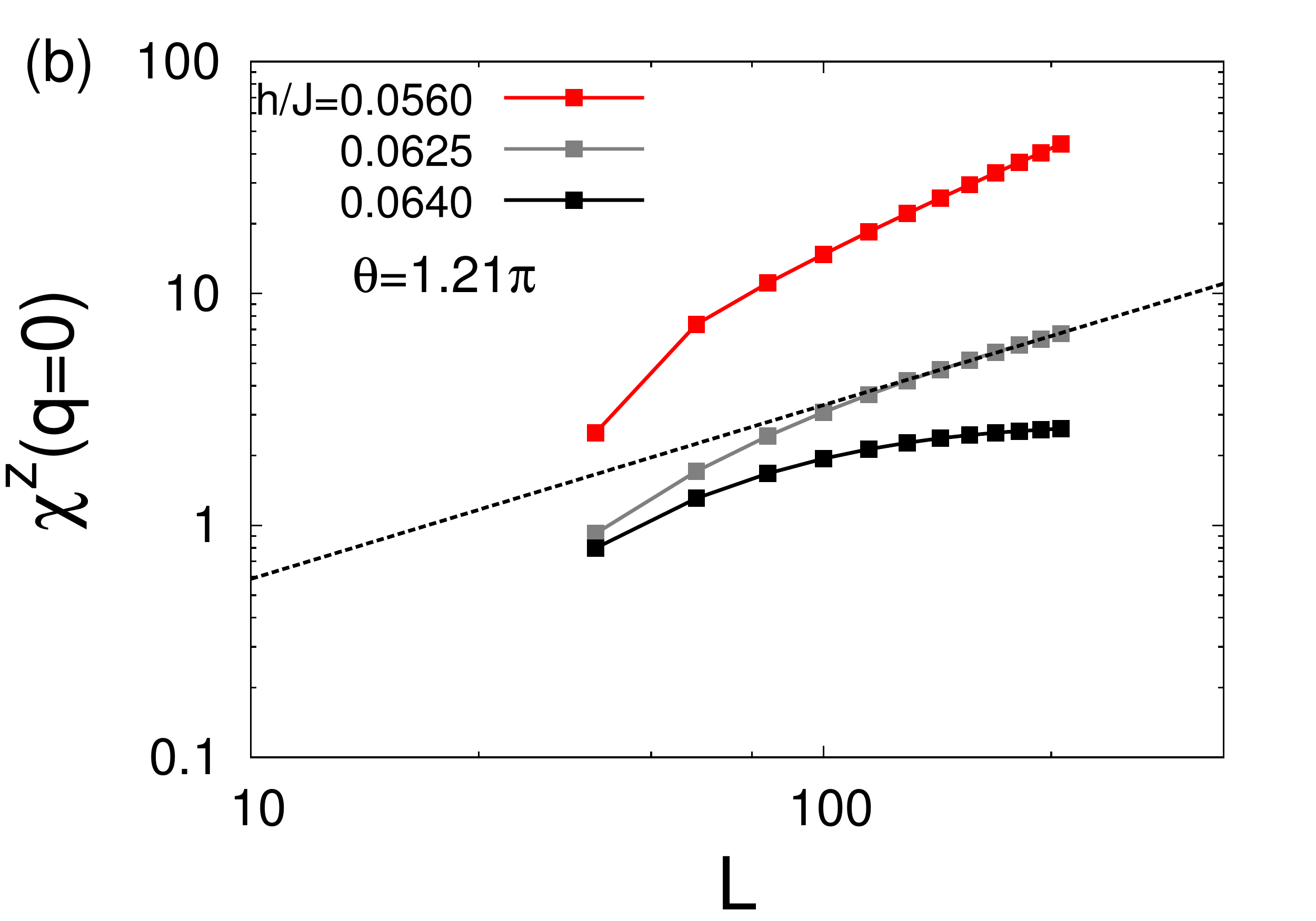}
\end{minipage}
\caption{Spin susceptibility as a function of system size $L$ and magnetic field $h$ for $\theta=0.6\pi$ (a) and $\theta=1.21\pi$ (b). In the SFM phase $\chi^z(0)\sim L$ (red) and in the SPM $\chi^z(0)\sim \mathrm{const.}$ (black). At the critical point the spin susceptibility diverges with $L$ like $\chi^z(0)\sim L^{\gamma/\nu}$ (grey) and using the value of $\nu$ found in Fig.~\ref{fig:xiL}, we find $\gamma=0.52(4)$ and  $0.50(4)$ for $\theta=0.6\pi$ and $1.21\pi$ respectively. } 
\label{fig:Sz_L_h}
\end{figure}

As shown in Fig.~\ref{fig:Sz_L_h}, in the SFM phase $\chi^z(q=0)\sim L$ and in the SPM phase $\chi^z(q=0)\sim$ constant in the large $L$ limit, whereas $\chi^\perp(q=\pm\eta)\sim L$ for $h\neq0$ with a slope that grows with $h$ (not shown). 
At the critical point the longitudinal spin susceptibility diverges like $\chi^z(q=0)\sim |\delta|^{-\gamma}$ and for finite $L$ this becomes rounded out like $\chi^z(q=0)\sim L^{\gamma/\nu}$. 
At $h=h_c$ we provide estimates of $\gamma/\nu$ for $\theta =0.6\pi$ and $1.21\pi$ as shown in Fig.~\ref{fig:Sz_L_h}.
We find for $\theta=0.6\pi$, $\gamma/\nu = 0.77$ and for $\theta=1.21\pi$, $\gamma/\nu = 0.75$. Using our estimates of $\nu$ we find $\gamma=0.52(4)$ for $\theta=0.6 \pi$ and $\gamma=0.50(4)$ for $\theta = 1.21 \pi$ and conclude $\gamma \approx 1/2$.

Our estimate of the correlation length exponent $\nu$ lies in the range that could be consistent with one of three different (known) universality classes: Perturbative renormalizaiton group (RG) calculations on  
the classical Ising model with strong long-range dipolar interactions~\cite{Larkin-1969,ZinJustin,Aharony-1973}
and the classical XY chiral spin liquid transition~\cite{Dimitrova-2014}, find for the thermal transition in $d=3-\epsilon$ dimensions at two loop order~\cite{ZinJustin,Dimitrova-2014} $\nu\approx0.626$
for $\epsilon=1$. However, the three-dimensional Ising universality class also has $\nu=0.629971(4)$ Ref.~\onlinecite{Poland-2016} and the three-dimensional XY universality class has $\nu=0.67155(27)$ Ref.~\onlinecite{Campostrini-2001} and our results cannot distinguish between the three.  However, our estimate of $\gamma$ does not agree with any of these  universality classes. While this is not conclusive, it does suggest that the universality class we have discovered is novel. It is also possible that we need to compare different channels, as
it is non-trivial to compare our estimate of $\gamma$ to those of the RG since we are computing the order parameter susceptibility from $\langle S^z(r)S^z(r')\rangle$, whereas the RG in Ref.~\cite{Dimitrova-2014} computes a ``twist'' susceptibility, which corresponds to a correlation function quartic in spin operators.
At this stage, we are inclined to believe that our system indeed belongs to a new universality class not studied before in the literature, but more work (and possibly more accurate numerical estimates of critical exponents) are necessary before any definitive conclusion can be reached.

For the DMRG calculations in the SFM using a bond dimension of $M=50$ we find a truncation error at most on the order of $\sim10^{-12}$, while close to the transition to the SPM phase the truncation error increases slightly to $\sim 10^{-11}$. In both the SFM and SPM phases the results have a very weak dependence on $M$ and increasing $M$ from $50$ to $100$ only effects the ground state energy near the transition on the order of $4\cdot10^{-5}$, while in the SFM phase the results agree to within numerical precision. Therefore, for all of the results presented in this manuscript for the SFM phase we use $M=50$ unless stated otherwise. In the calculations that follow we use an infinite-DMRG initialization procedure~\cite{De-2008} followed by 5 finite size DMRG sweeps until the data is well converged (unless otherwise specified).

\subsection{Classical Solution}
We find that a relatively small bond dimension is required to capture the SFM ground state in DMRG, which suggests we can describe the physics of this phase (at least close to $\theta=\pi$) with a classical product state approximation. In this subsection, we solve the model in Eq.~(\ref{eqn:spin_model}) in the large-$S$ limit of the model as a classical energy functional of vector degrees of freedom ${\bf S}_i \rightarrow (\cos \varphi_i \sin\vartheta_i,\sin \varphi_i \sin\vartheta_i,\cos\vartheta_i)$ with unit norm and angles $\vartheta_i$ and $\varphi_i$. The energy is then variationally minimized, giving static ground state spin configurations which can be compared with the DMRG results.

\begin{figure}[t]
\centering
\begin{minipage}{.5\textwidth}
  \centering
  \includegraphics[width=0.95\linewidth]{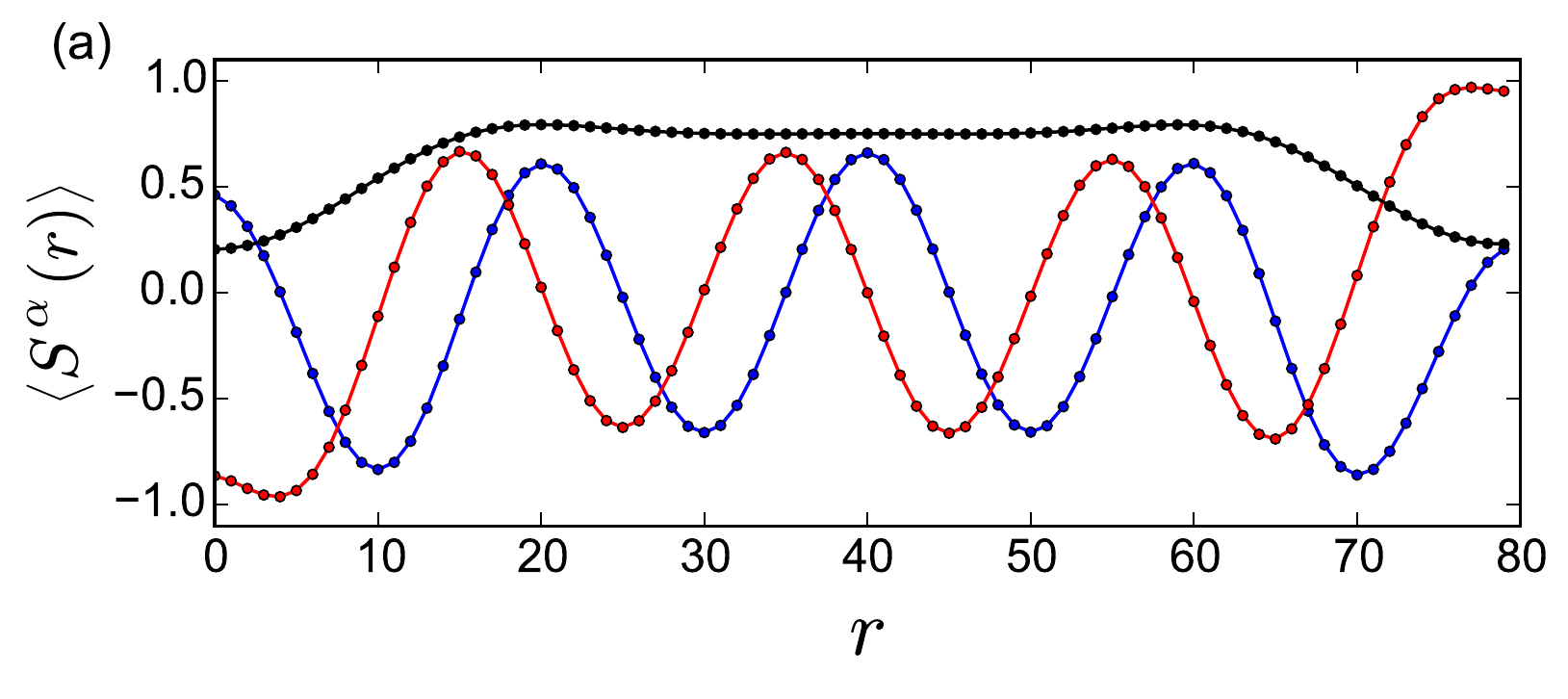}
  \includegraphics[width=0.95\linewidth]{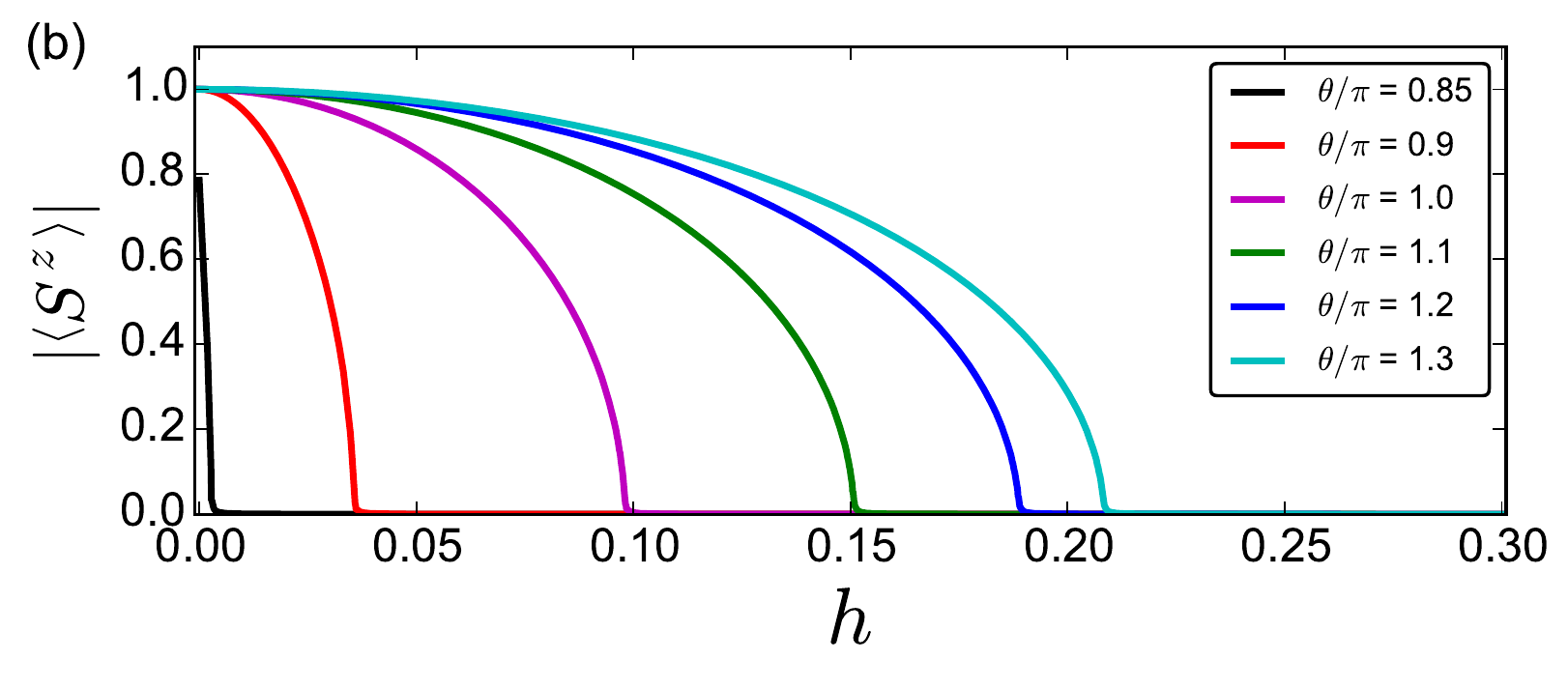}
  \includegraphics[width=0.95\linewidth]{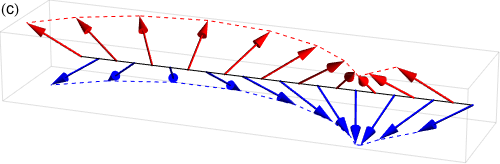}
\end{minipage}
\caption{Classical approximation results in the large-$S$ limit. In (a) we show a typical spin configuration corresponding to the SFM ($\theta = 1.1\pi$, $h = 0.1J$)  In (b) we plot the order parameter of the SFM phase with increasing $h$, demonstrating the existence of a classical SFM to SPM transition, and the effect of changing $\theta$. In (c) we show a 3D plot of a segment of the staggered phase, demonstrating that neighboring spins are nearly perpendicular (with an overall spiral component), resulting from a purely classical frustration due to a positive $K$ ($\theta = 0.7\pi$, $h=0.1J$). This phase does not appear in the DMRG results, suggesting that quantum effects preserve the stability of the ferromagnet for positive $K$.}

\label{fig:classical_panel}
\end{figure}

In Fig.~\ref{fig:classical_panel}(a), we show a typical spin configuration in the SFM phase, with open boundary conditions. As with the DMRG results, we also observe here a boundary effect classically, while toward the center of the chain $S^z(r)$ is nearly uniform with $(S^x(r), S^y(r)) \propto (\cos \eta r, \sin \eta r)$.

 \begin{figure}[t]
\centering
\begin{minipage}{.5\textwidth}
  \centering
  \includegraphics[width=0.7\linewidth,angle=-90]{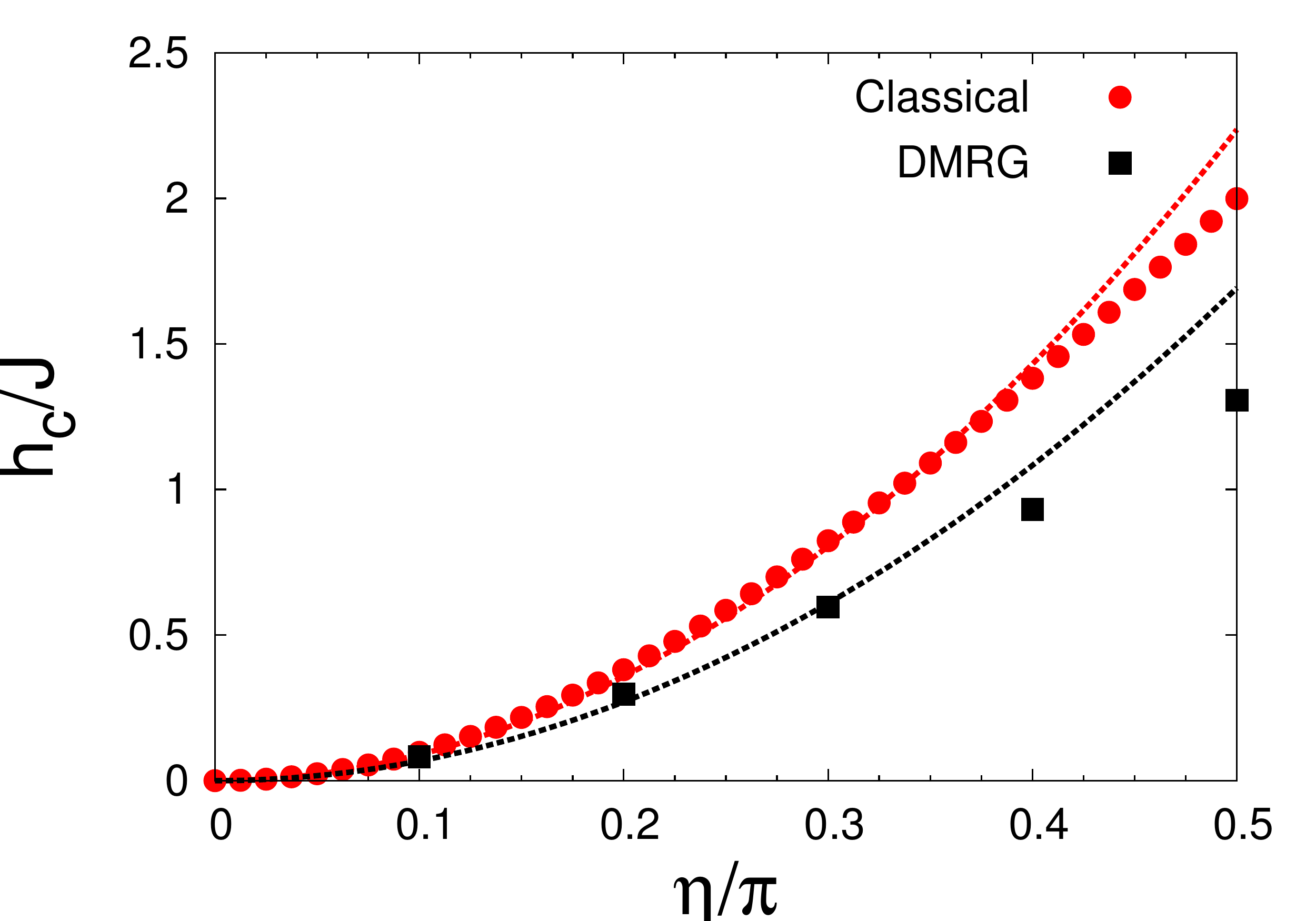}
\end{minipage}%
\caption{Critical field $h_c$ as a function of the pitch of the spiral $\eta$ for $\theta=0$, for DMRG (black squares) and the classical solution (red circles), the dashed lines are a fit to the mean field result $h_c \propto \eta^2$. We find the deviation between the quantum and classical solutions grow with $\eta$ and the fit to the mean field result only works well for the DMRG and classical data at small $\eta$. Note that $h_c(\eta=0.1 \pi)$ from DMRG and the classical solution do deviate (see Fig.~\ref{fig:phase_diagram} ).} 
\label{fig:eta_vs_hc}
\end{figure}

The classical SFM has a transition to a classical SPM with increasing $h$, shown in Fig.~\ref{fig:classical_panel}(b) for fixed $\eta=0.1\pi$ and several values of $\theta$. For $\theta$ corresponding to negative $K$, the SFM becomes increasingly robust against the SOC field, as both $\tilde{J}$ and $\tilde{K}$ terms try to maximize $({\bf S}_i \cdot {\bf S}_{i+1})$. However, for positive $\tilde{K}$ the classical SFM is frustrated since $\tilde{J}$ is trying to align neighboring spins while $\tilde{K}$ is trying minimize $|{\bf S}_i \cdot {\bf S}_{i+1}|$. This eventually gives way to a phase with a \emph{staggered} component along $S^z$, such that neighboring spins are nearly orthogonal to satisfy the large $\tilde{K}$. We illustrate this phase with a 3D plot of the spins in Fig.~\ref{fig:classical_panel} (c).
This staggered phase does not appear in the DMRG and is purely an artifact of the classical approximation. Thus, we find the classical approximation qualitatively agrees with the DMRG results very close to $\theta = \pi$, which is natural since the the ferromagnet with $h=0$ is a product state and minimally entangled.

Away from this special point, however, the classical approximation fails in capturing either the dimer phase (discussed below) for negative $K$ \emph{or} the observed persistence of the quantum SFM to large positive $K$. In some sense, we can regard the classically-unexpected stability of the SFM phase in the quantum problem as resulting from an order by quantum disorder phenomenon, where quantum fluctuations are able to overcome the classical frustration due to a large positive $\tilde{K}$.

\subsection{Landau Theory}
\label{subsec:LT}
In this subsection we provide a phenomenological theory of the spiral ferromagnetic phase. We start with the assumption of an $O(3)$ quantum $\phi^4$ theory where $\bm{\phi}(r,\tau)$ is a vector that captures the magnetic fluctuations in the ordered phase, $\tau$ denotes imaginary time, and in the absence of the spiral magnetic field the low energy spin excitations disperse quadratically, i.e. $\epsilon(q)\sim q^2$. Using these assumptions we write down a phenomenological Landau theory for the ordered spiral ferromagnetic phase with an action
\begin{eqnarray}
S &=&\int_0^L \,dr\int_0^{\beta}\, d\tau\Big(\bm{\phi}\cdot\partial_{\tau}\bm{\phi} + K(\partial_r \bm{\phi})^2 + r_0 \bm{\phi}^2 
\\
&+& u_0 \bm{\phi}^4-h\cos(\eta r)\phi^x+h\sin(\eta r)\phi^y \Big).
\end{eqnarray}
where $K$ is the stiffness, $r_0(<0)$ is the mass, $u_0$ is the interaction between the collective modes, and the field is coupled to the spiral magnetic field.
We begin by determining the mean field solution using the ansatz $\bm{\phi}_{\mathrm{MF}} = -\phi_{\perp,0}( \cos(q r)\hat{x}- \sin(q r)\hat{y}) +M^z_0 \hat{z}$, this yields
\begin{eqnarray}
q&=&\eta, 
\\ 
\phi_{\perp,0} &=& \frac{h}{2K\eta^2},
\\
M_0^z &=& \pm\sqrt{\frac{-r_0}{2u_0} - \phi_{\perp,0}^2} = \pm\frac{\sqrt{h_c^2-h^2}}{2 K \eta^2}
\end{eqnarray}
and we find the critical field 
\begin{equation}
h_c = K\eta^2\sqrt{-2r_0/u_0}.
\end{equation}
\begin{figure}[t]
\centering
\begin{minipage}{.25\textwidth}
  \centering
  \includegraphics[width=0.7\linewidth,angle=-90]{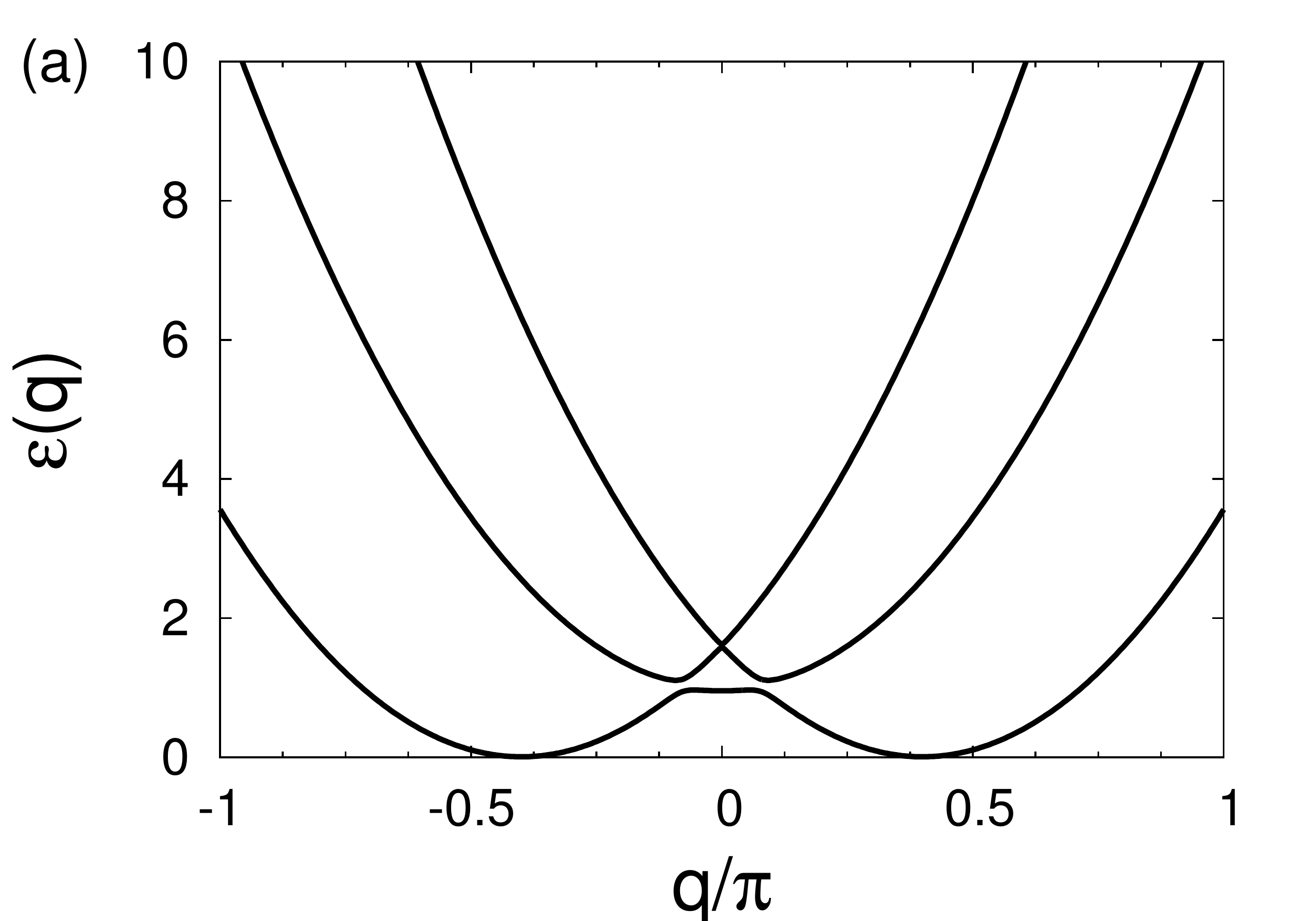}
\end{minipage}%
\begin{minipage}{.25\textwidth}
  \centering
  \includegraphics[width=0.7\linewidth,angle=-90]{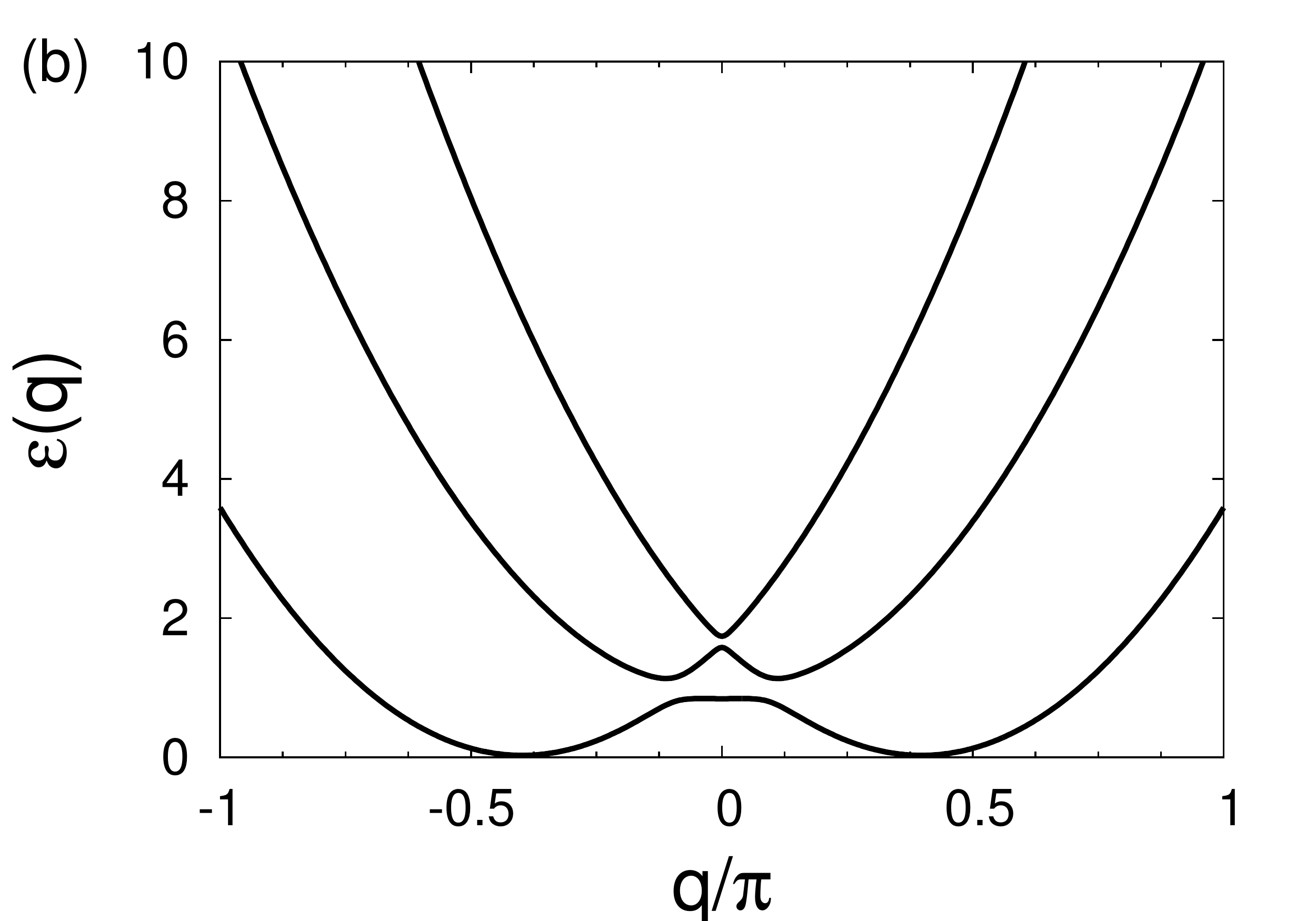}
\end{minipage}
\newline
\begin{minipage}{.25\textwidth}
  \centering
  \includegraphics[width=0.7\linewidth,angle=-90]{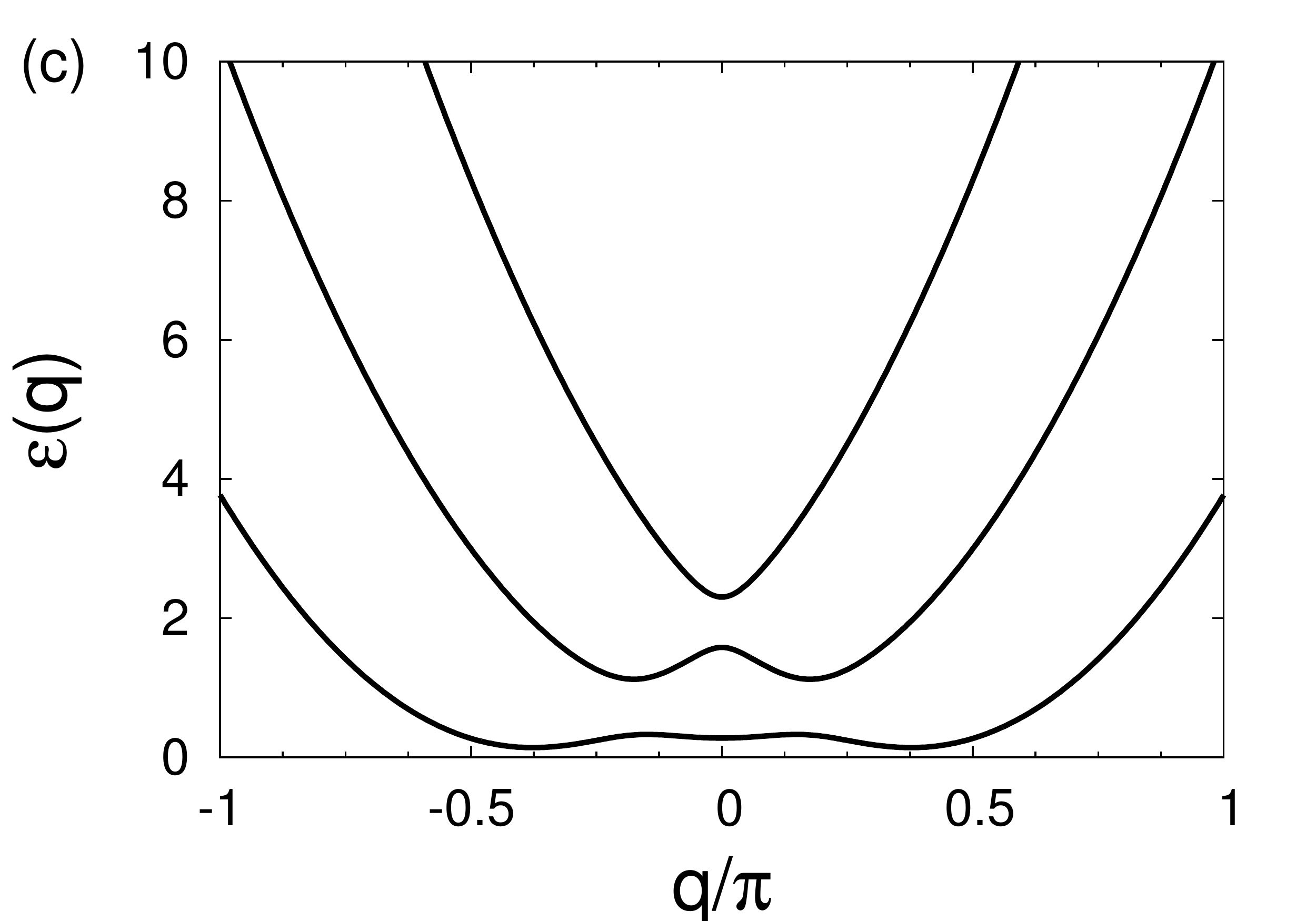}
\end{minipage}%
\begin{minipage}{.25\textwidth}
  \centering
  \includegraphics[width=0.7\linewidth,angle=-90]{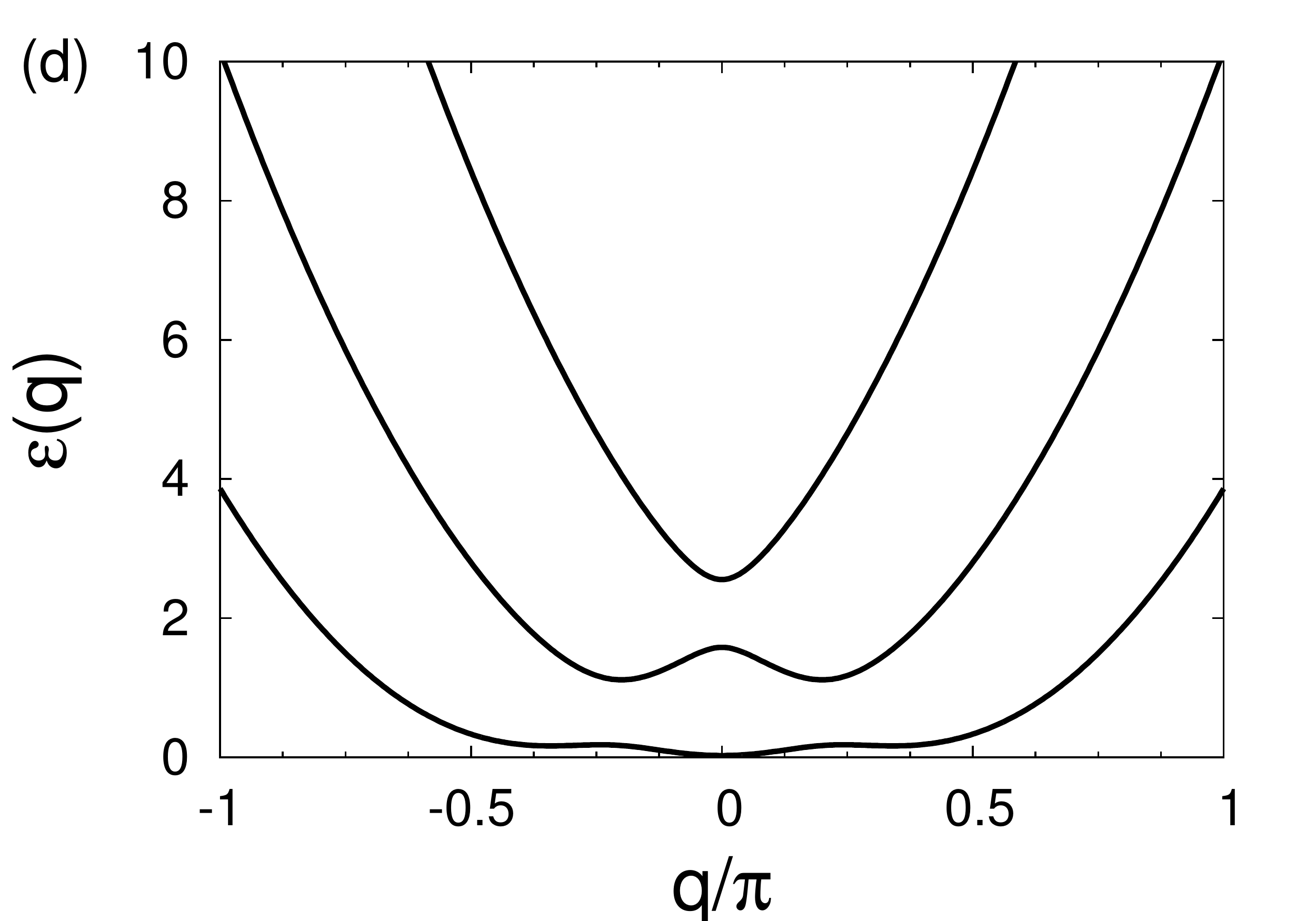}
\end{minipage}
\caption{Dispersion from Landau theory truncating the effective action in Eq.~(\ref{eqn:lt}) to quadratic order. As a prototypical example, we take $r_0/K=-0.5$, $u_0/K=0.2$, and $\eta=0.4\pi$, which has a critical field $h_c/K=3.53$. Evolution of the spin wave dispersion as a function of increasing magnetic field $h/K=0.5$ (a), $h/K=1.0$ (b), $h/K=2.5$ (c) and $h/K=3.0$ (d). } 
\label{fig:Eq}
\end{figure}
The mean field solution is in excellent agreement with both the DMRG results  and the classical solution for $\langle {\bf S}_i \rangle$ [see Figs.~\ref{fig:mx} and~\ref{fig:classical_panel} (a) and (b)]. In Fig.~\ref{fig:eta_vs_hc} we show the comparison of $h_c$ versus $\eta$ as computed from DMRG and the classical model, interestingly we find the data for both follows the prediction of Landau theory at small $\eta$.

We now expand about the ordered state via $\bm{\phi}=\bm{\phi}_{\mathrm{MF}}+\bm{\psi}$, under the assumption that the fluctuations $\bm{\psi}$ are small compare to the mean field solution.  We find it is convenient to rotate the fluctuations at each site about the $z$ axis by the pitch of the magnetic field $\eta$, and therefore work with the rotated degrees of freedom 
\begin{eqnarray}
\left(
  {\begin{array}{c}
  \tilde{\psi}_x \\
  \tilde{\psi}_y
  \end{array} } 
  \right)=\left( {\begin{array}{cc}
  \cos(\eta r) & -\sin(\eta r) \\
   \sin(\eta r) &  \cos(\eta r)
  \end{array} } \right)
  \left(
  {\begin{array}{c}
  \psi_x \\
  \psi_y
  \end{array} } 
  \right)
\end{eqnarray}
%%(i.e. $U(\eta r)$ applies a rotation of $\eta r$ at the location $r$ about $\hat{z}$). 
%Here $\bm{\psi}_{\perp}^T=(\psi^x,\psi^y)$ 
and $\tp^z=\psi^z$. To $O(\tilde{\psi}^4)$ this yields
\begin{eqnarray}
S &=& \frac{\beta L}{2}\left(r_0\bm{\phi}_{\mathrm{MF}}^2-h\phi_{\perp,0} \right)+\int_0^L \,dr\int_0^{\beta}\, d\tau\mathcal{L}_{\psi},
\nonumber
\\
\mathcal{L}_{\psi}
&=&\btp\cdot\partial_{\tau}\btp +K \vert(\partial_r +i\eta \sigma_y)\btp_{\perp}\vert^2+K(\partial_r \tilde{\psi}^z)^2
\nonumber
\\
&+&u_0 \left(   2 (M_0^z \tilde{\psi}^z - \phi_{\perp,0} \tilde{\psi}^x)+\vert \btp \vert^2 \right)^2.
\label{eqn:lt}
\end{eqnarray}
In the rotated frame the spiral magnetic field appears as a gauge field acting on $\btp_{\perp}$ (i.e. $\partial_r \rightarrow \partial_r + i\sigma_y \eta$, where $\sigma_y$ is the $y$ Pauli matrix), which shifts the zero of the spin wave dispersion. 
In addition to the standard quartic interaction, the effective action we have derived also has anisotropic cubic interactions induced by the SOC. At $h_c$, $M_0^z\rightarrow 0$ and the assumption that the fluctuations about the mean field solution are small no longer apply. 

%As we have already identified the universality class of the transition numerically we do not seek to do so here. Instead 
We use the Landau theory to determine the low energy spin-wave excitations in the SFM. We therefore restrict ourselves to the action at the quadratic level and drop fluctuations on the order of $O(\psi^3)$ and higher. We Fourier transform the action to momentum space $(q)$ and diagonalize the $3\times3$ matrix at each $q$ to determine the spin wave excitations for $h<h_c$. As shown in Fig.~\ref{fig:Eq}, we find that gapless spin waves persist in the presence of the field and disperse quadratically about $\pm \eta$ for $h \ll h_c$. For increasing $h$ we find the spin wave modes eventually become gapped out and the minimum shifts to $q=0$.

\section{Spiral Dimer phase}
\label{sec:sd}

We now come to the dimerized regime of the phase diagram, where the ground state can spontaneously break the translational symmetry in the model.
In the interval $1.25\pi \lesssim \theta \le 3\pi/2$ with $h=0$ the ground state is spontaneously dimerized (breaking the translational symmetry of the lattice) and gapped. Turning on a weak magnetic field, much smaller then the gap, therefore should not destroy the non-zero dimer order parameter. Near $\theta=1.25\pi$ this requires an extremely small field since the gap is exponentially small in $|\theta-1.25\pi|$ (see Refs.~\onlinecite{Hu-2014,Gerster-2014,Matteo-2005,Lauchli-2006}).   Instead, it couples neighboring valence bonds, introducing significant entanglement in the ground state, and as a result we need a much larger bond dimension than in the SFM phase. 
It is therefore no longer reasonable to use a small bond dimension like we did above (e.g., $M=50$ gives a large truncation error of the order of $10^{-4}$). However, we find that $M=200$ is sufficient to limit the truncation below a reasonable $10^{-8}$ and keep the computational costs at an acceptable level. 
%(without good quantum numbers to be preserved, each run with $L=120$ took already about 1 week). 
We present results accordingly, unless otherwise stated.

 \begin{figure}[t]
\centering
\begin{minipage}{.25\textwidth}
  \centering
  \includegraphics[width=0.7\linewidth,angle=-90]{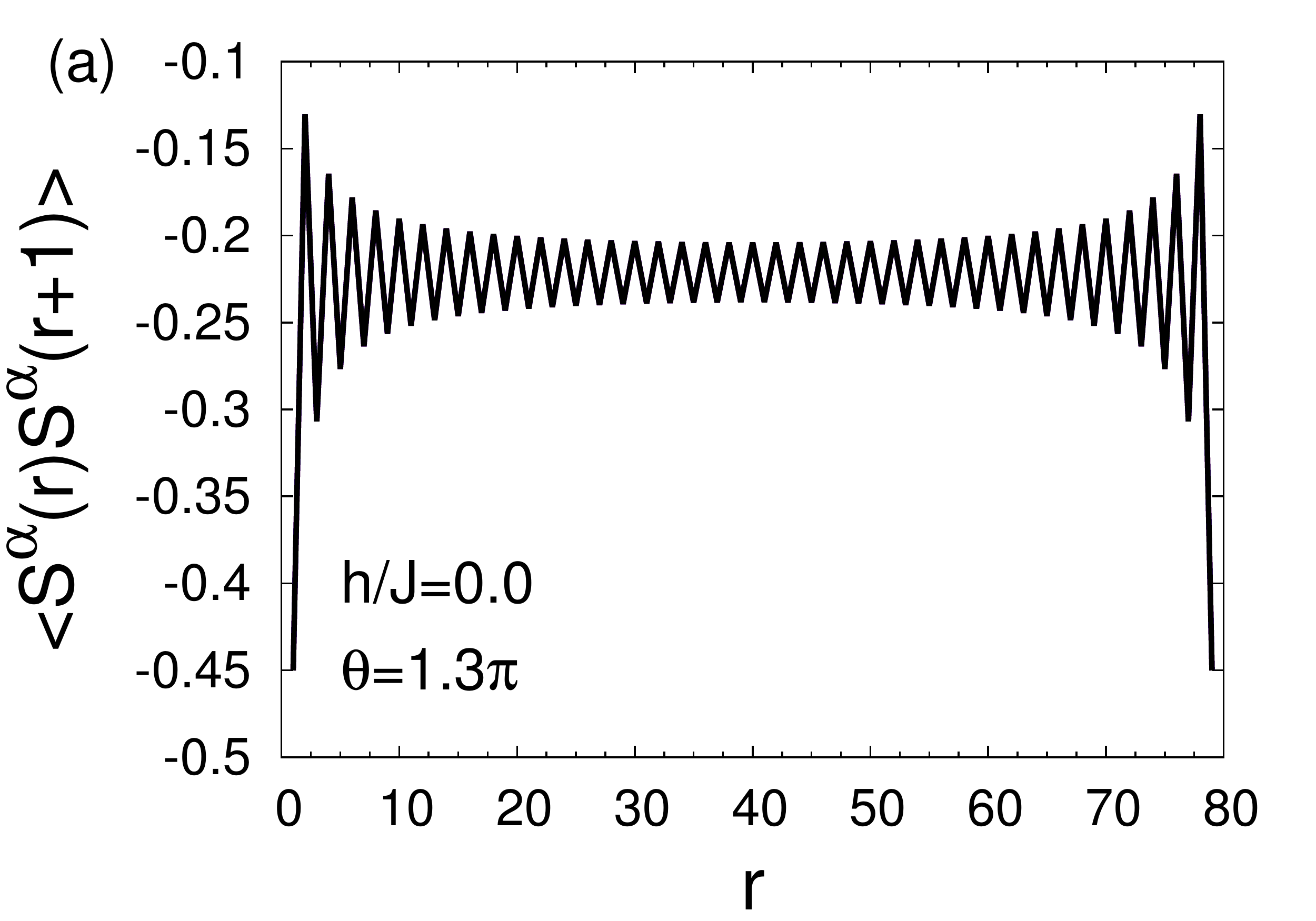}
\end{minipage}%
\begin{minipage}{.25\textwidth}
  \centering
  \includegraphics[width=0.7\linewidth,angle=-90]{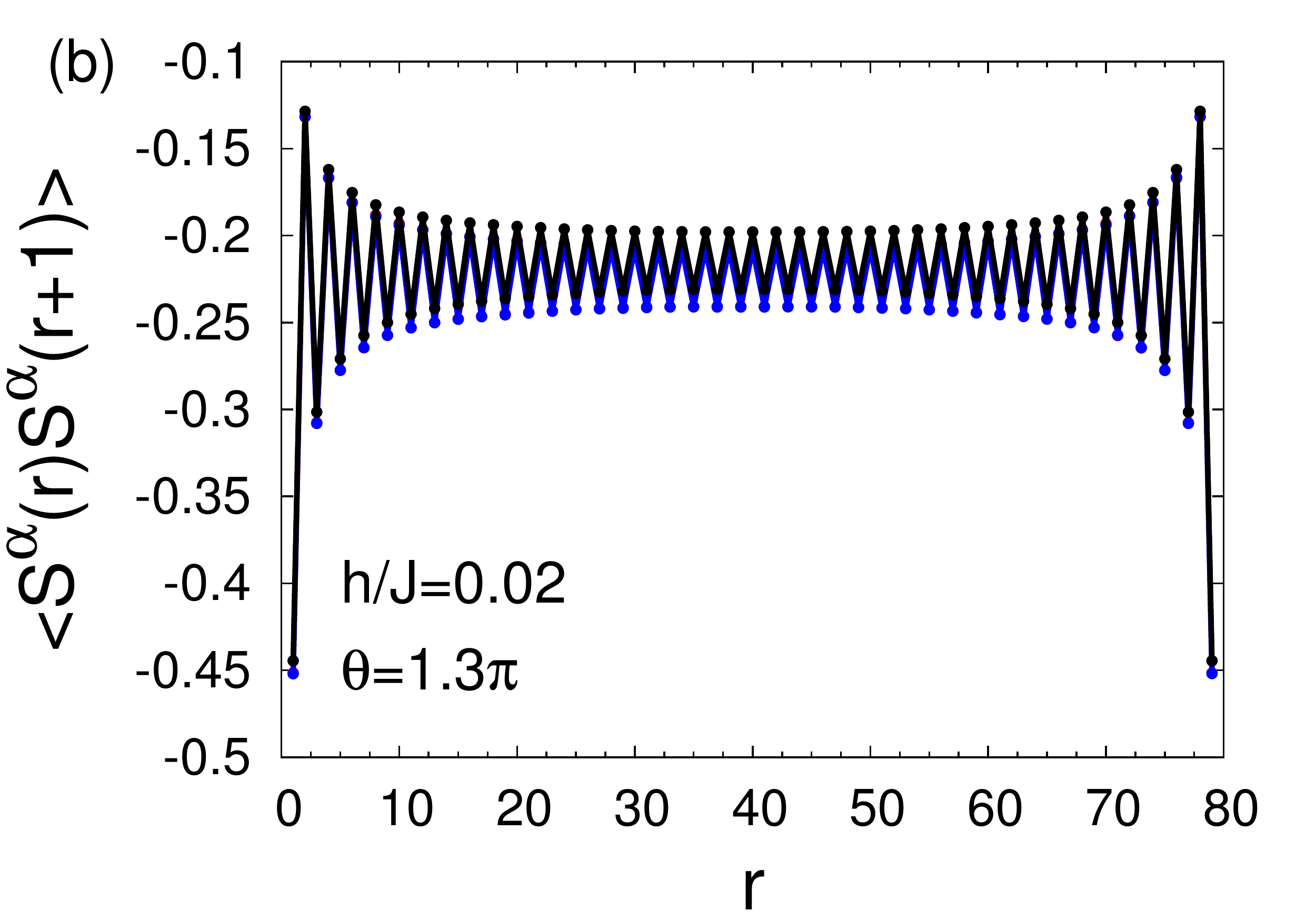}
\end{minipage}
\newline
\centering
\begin{minipage}{.25\textwidth}
  \centering
  \includegraphics[width=0.7\linewidth,angle=-90]{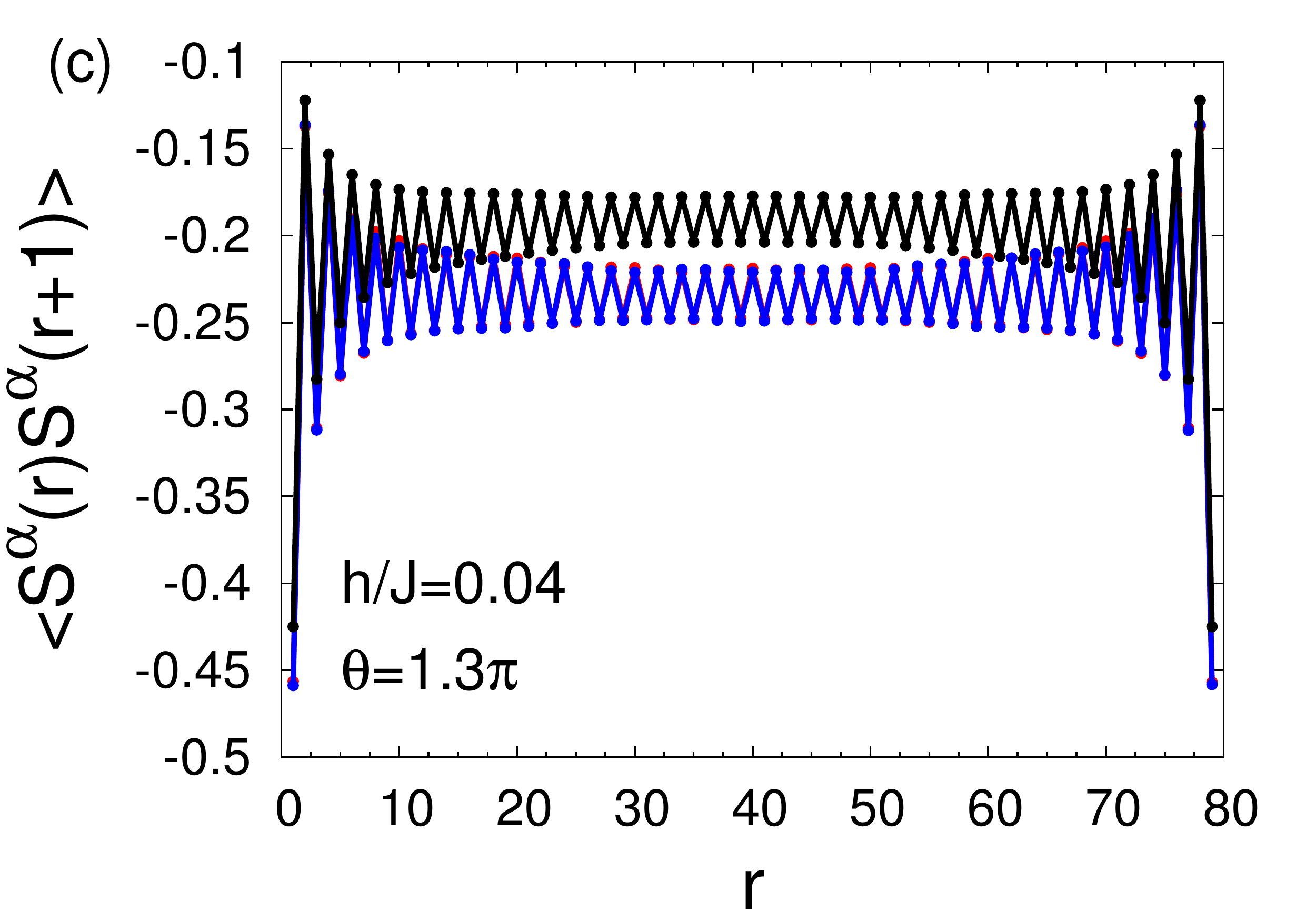}
\end{minipage}%
\begin{minipage}{.25\textwidth}
  \centering
  \includegraphics[width=0.7\linewidth,angle=-90]{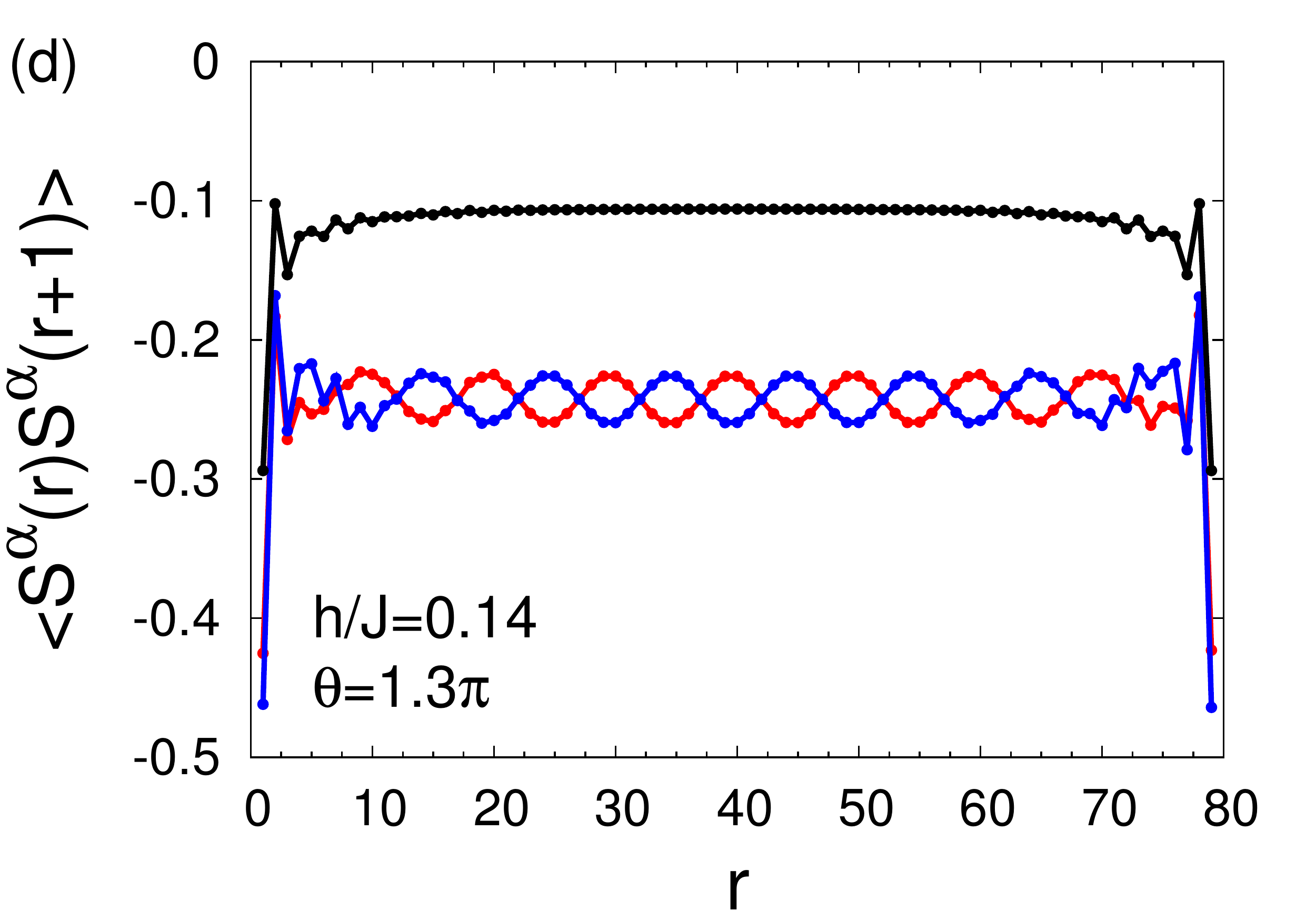}
\end{minipage}
\newline
\centering
\begin{minipage}{.25\textwidth}
  \centering
  \includegraphics[width=0.7\linewidth,angle=-90]{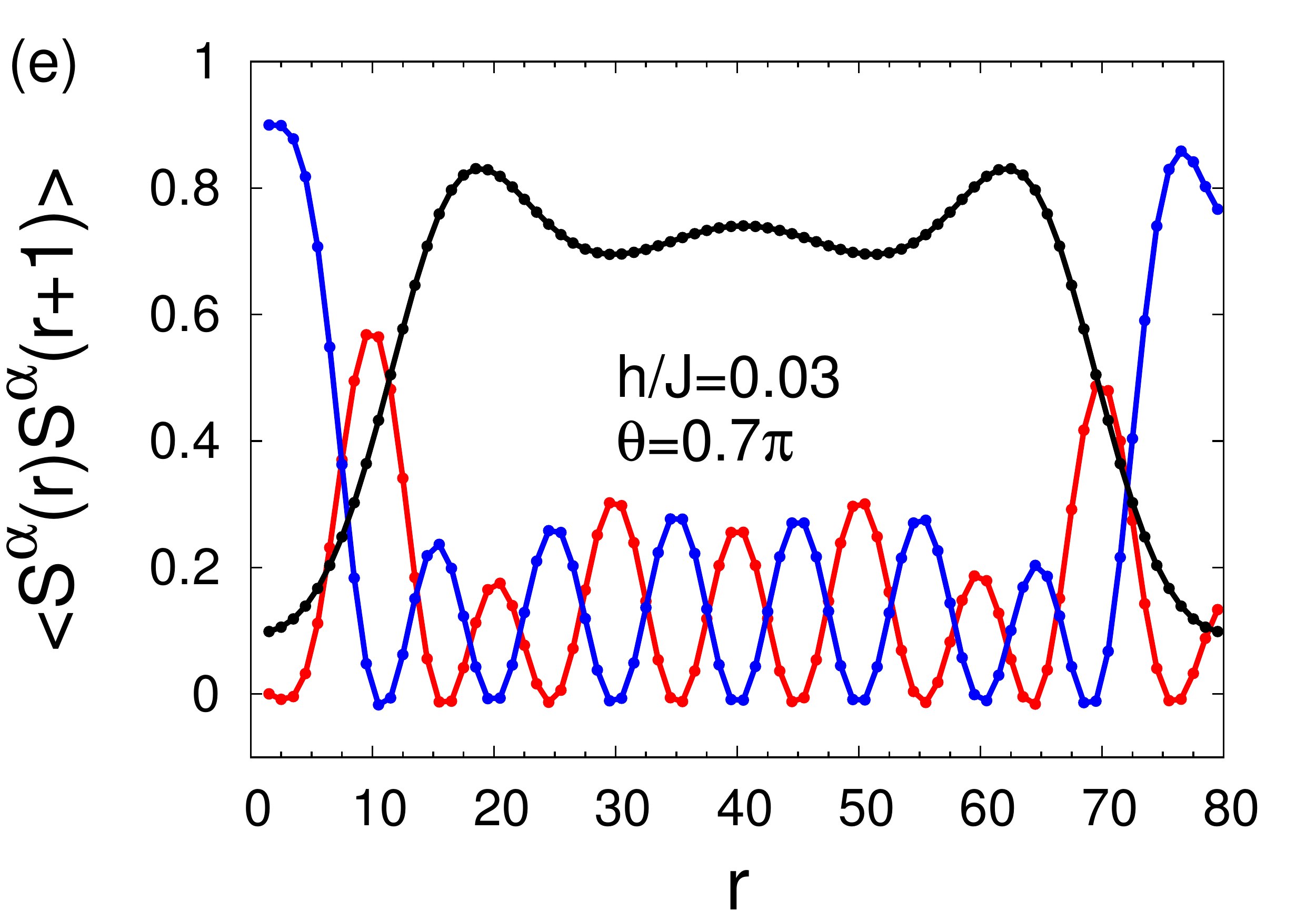}
\end{minipage}%
\begin{minipage}{.25\textwidth}
  \centering
  \includegraphics[width=0.7\linewidth,angle=-90]{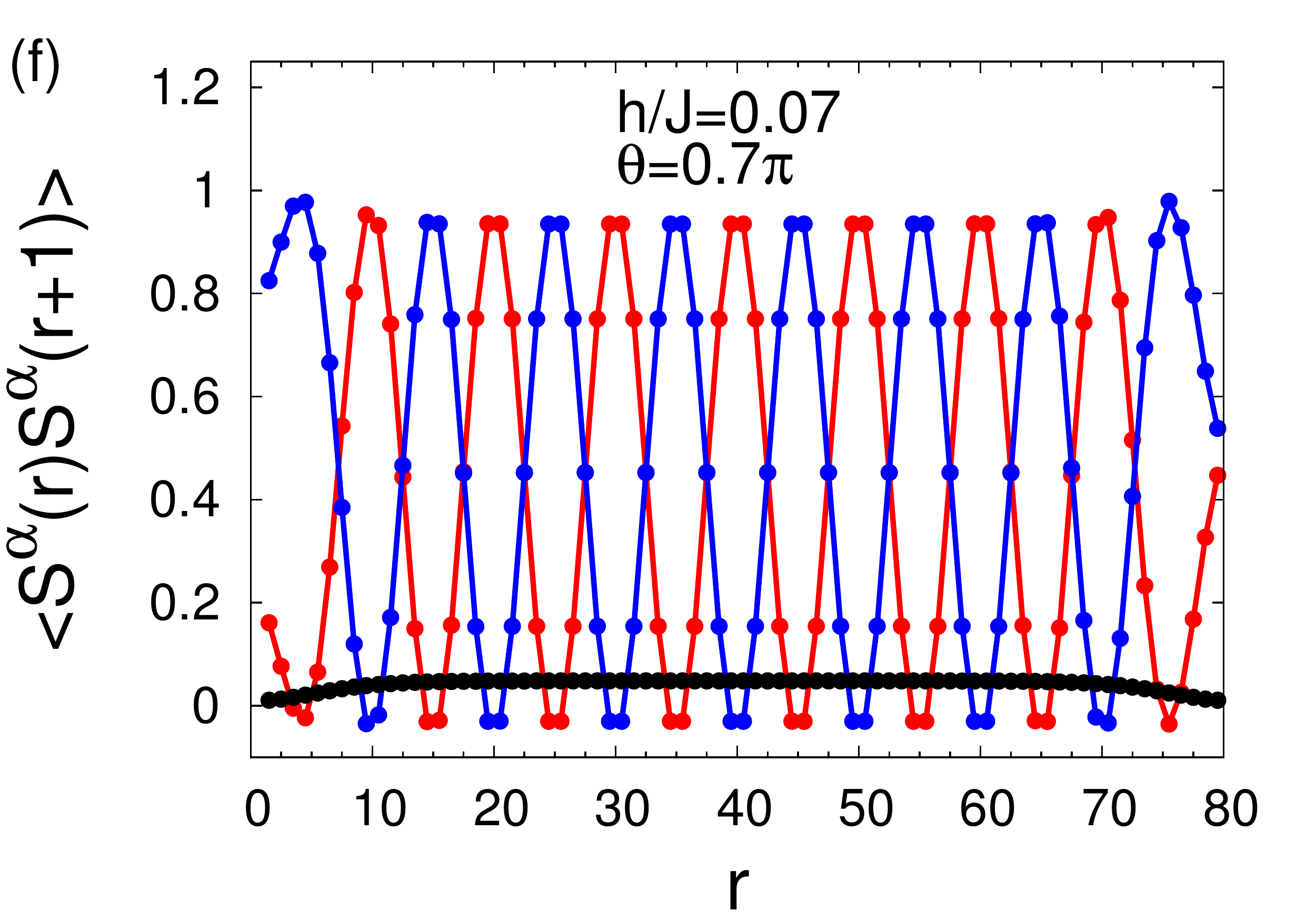}
\end{minipage}
\caption{Average nearest neighbor spin product along the chain for $\theta=1.3\pi$ and $L=80$ in the SD phase (a), (b), (c), in the SPM phase for $\theta >1.25\pi$ (d). In contrast, for $\theta < 1.25 \pi$ (e) is in the SFM phase and (f) is in the SPM phase. We find the effect of the magnetic field in the SD phase is to break the spin symmetry between the $x$ (red), $y$ (blue), and $z$ (black) components of the spin. For large fields in the SPM phase the translational symmetry is restored and the value of $\theta$ dictates the sign of the spin product.} 
\label{fig:SSD_L_h}
\end{figure}

 \begin{figure}[t]
\centering
\begin{minipage}{.25\textwidth}
  \centering
  \includegraphics[width=0.7\linewidth,angle=-90]{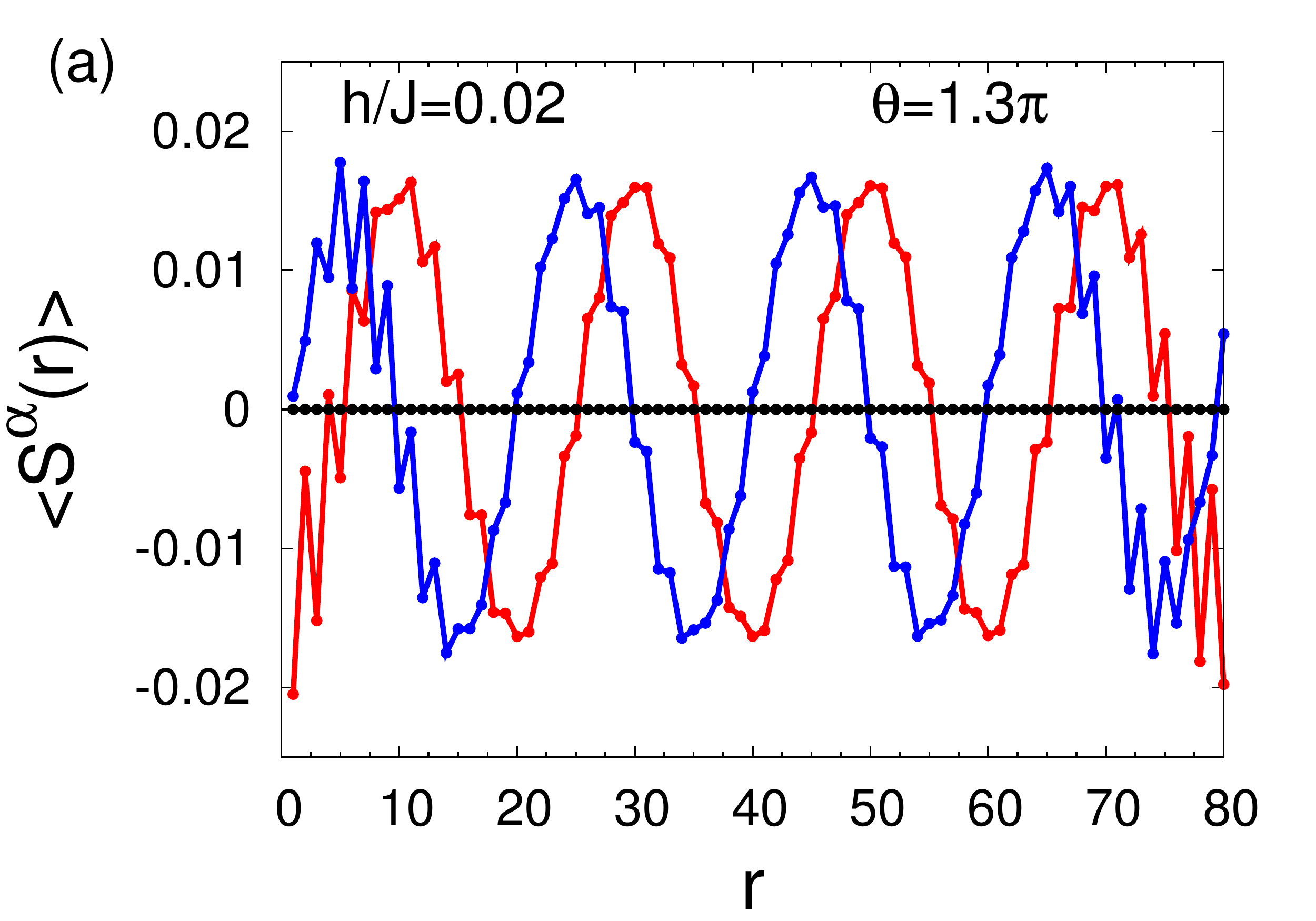}
\end{minipage}%
\begin{minipage}{.25\textwidth}
  \centering
  \includegraphics[width=0.7\linewidth,angle=-90]{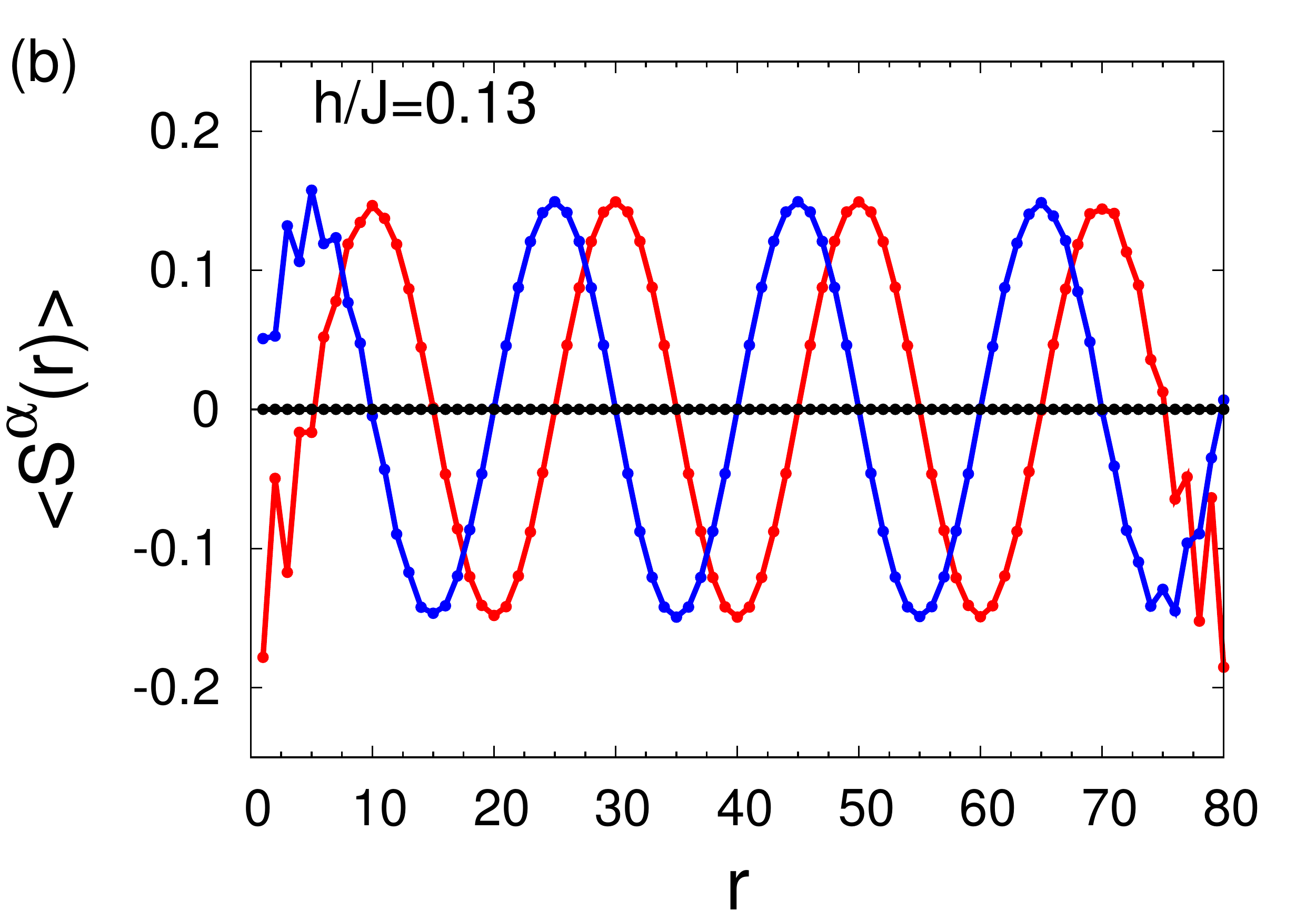}
\end{minipage}
\caption{Average spin along the chain for $\theta=1.3\pi$ in the SD phase (a) and in the SPM phase (b) for $L=80$ and $M=200$ and $x$ (red), $y$ (blue), and $z$ (black) spin components. The dimerization is visible in the average spin in the SD phase, whereas for large fields the translational symmetry is restored in the SPM phase.} 
\label{fig:SD_L_h}
\end{figure}

Introducing the spiral magnetic field has enlarged the unit cell from one site to $2\pi/\eta$ (in units of the lattice spacing), and we have translational symmetry across these unit cells for $\theta<1.25\pi$ as we have established in Sec.~\ref{sec:sf}A. In order to study the ground state breaking the translational symmetry of the model, 
we define the dimer order parameter as $D=| \langle d(r=L/2) \rangle|$, where
\begin{equation}
d(r) =  {\bf S}_{r-1}\cdot{\bf S}_{r}  -  {\bf S}_{r}\cdot{\bf S}_{r+1},
\label{eqn:dimer}
\end{equation}
which is non-zero (in the thermodynamic limit) when the translational symmetry of the model is broken. However, it is not a priori obvious whether this order parameter will be skewed by the presence of a spiral magnetic field, which has enlarged the unit cell to $2\pi/\eta$. Nonetheless, as shown in Fig.~\ref{fig:SSD_L_h} (a), (b), and (c) for fixed $L$, the dimer order parameter along the chain remains non-zero, and the effect of the magnetic field is to break the symmetry between the $x$, $y$, and $z$ components of $\langle S(r)^{\alpha} S(r+1)^{\alpha} \rangle$. As we increase the field, the dimerization is destroyed and the model eventually enters the SPM phase [see Fig.~\ref{fig:SSD_L_h} (d)]. 
One interesting contrast is that the behavior of this observable in the SFM phase mimics the spin while in the presence of dimerization these two are completely distinct [see Figs.~\ref{fig:mx},~\ref{fig:SSD_L_h} (e), and~\ref{fig:SD_L_h}]. 

As shown in Fig.~\ref{fig:SD_L_h}, despite the non-zero dimerization, the bare spin expectation values still follow the spiral magnetic field albeit with the broken translational symmetry ``imprinted'' upon the average spin. As a result, we find
\begin{eqnarray}
\langle S^x(r) \rangle &=& -A \cos(\eta r) + (-1)^r\Delta,
\\
\langle S^y(r) \rangle &=& A \sin(\eta r) + (-1)^r\Delta,
\end{eqnarray}
for an amplitude $A$, dimerization $\Delta$, and $\langle S^z(r) \rangle\approx 0$. Thus the spiral dimer phase (SD) is defined as having a non-zero dimer order parameter and dimerized spiral pattern in the average spin.

\begin{figure}[t]
\centering
\begin{minipage}{.25\textwidth}
  \centering
  \includegraphics[width=0.7\linewidth,angle=-90]{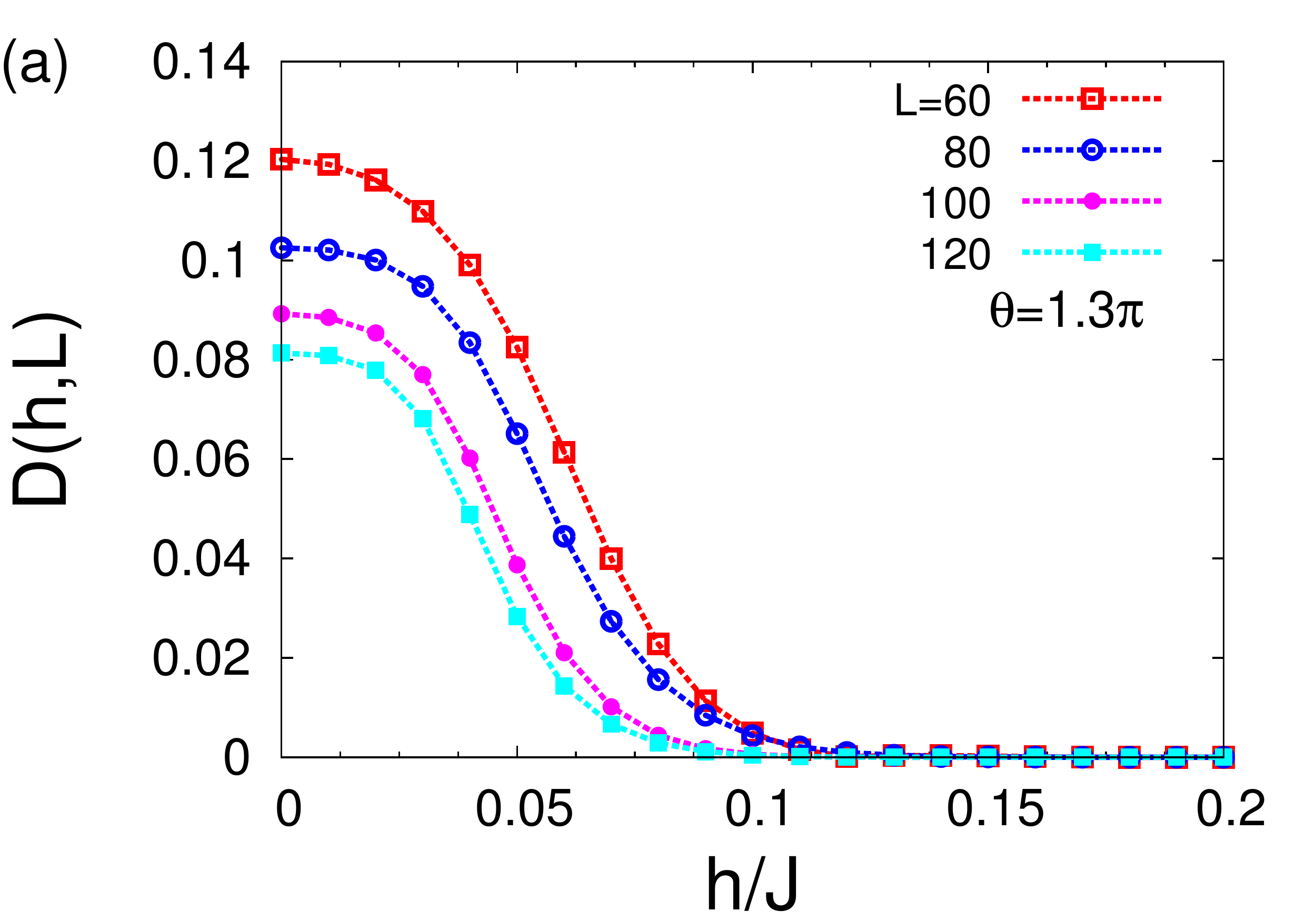}
\end{minipage}%
%\newline
\begin{minipage}{.25\textwidth}
  \centering
  \includegraphics[width=0.7\linewidth,angle=-90]{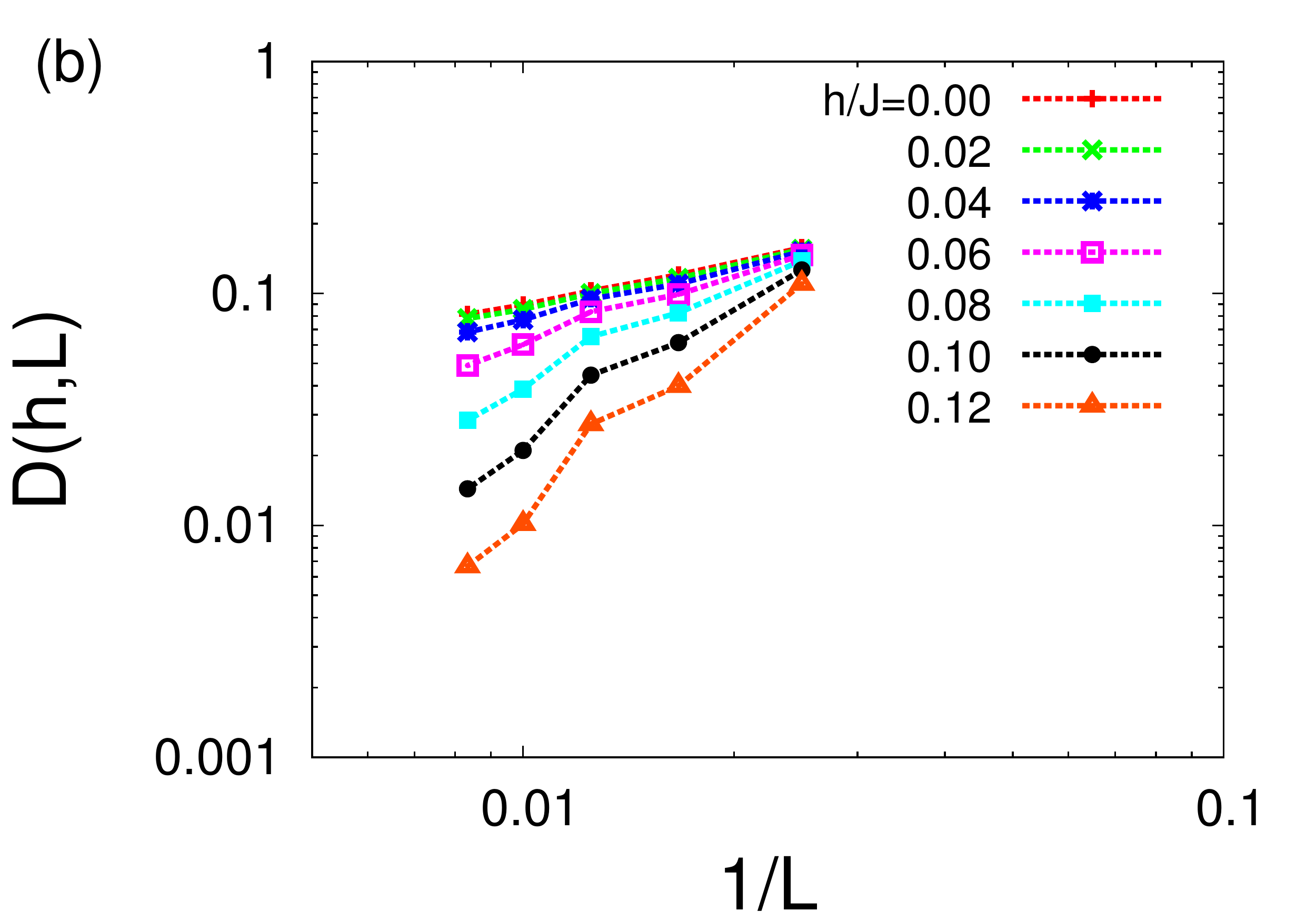}
\end{minipage}
\caption{Dimer order parameter defined in Eq. (\ref{eqn:dimer})  for $\theta=1.3\pi$ as a function of $h$ and $L$. $D$ versus $h$ for various $L$ (a) and $D$ versus $L$ for various $h$ (b). In the SD phase we find the dimer order parameter is saturating to a constant for large $L$ and weak $h$. At large $h$, $D$ goes to zero at large $L$, and we take an estimate for transition from the smallest value of $h$ where $D$ still seems to vanish at large $L$. } 
\label{fig:D_L_h}
\end{figure}
We now come to the finite size scaling of the dimer order parameter, for system sizes ranging from $L=40$ to $120$.
We find that the dimer order parameter is approaching a non-zero value in the large $L$ limit in the range $1.29\pi \lesssim \theta \le 3\pi/2$ despite the presence of a small spiral magnetic field. As shown in Fig.~\ref{fig:D_L_h} (a), for increasing $h$ we find that the dimer order parameter goes to zero, implying a continuous transition out of the dimerized phase into the SPM. 
Therefore, despite the presence of a non-zero spiral magnetic field, we still find that $D$ distinguishes the dimerized phase from the SFM and SPM phases in the large $L$ (or thermodynamic) limit.
However, as seen in Fig. ~\ref{fig:D_L_h} (b), finite size effects for the the dimer order parameter are not small. As a result $D$ is not particularly suited for a precise extraction of the critical phase boundary with the system sizes at our disposal. Therefore, we resort to an analysis of the dimer-dimer correlation function.

\begin{figure}[h!]
  \centering
\begin{minipage}{.44\textwidth}
  \includegraphics[width=0.65\linewidth,angle=-90]{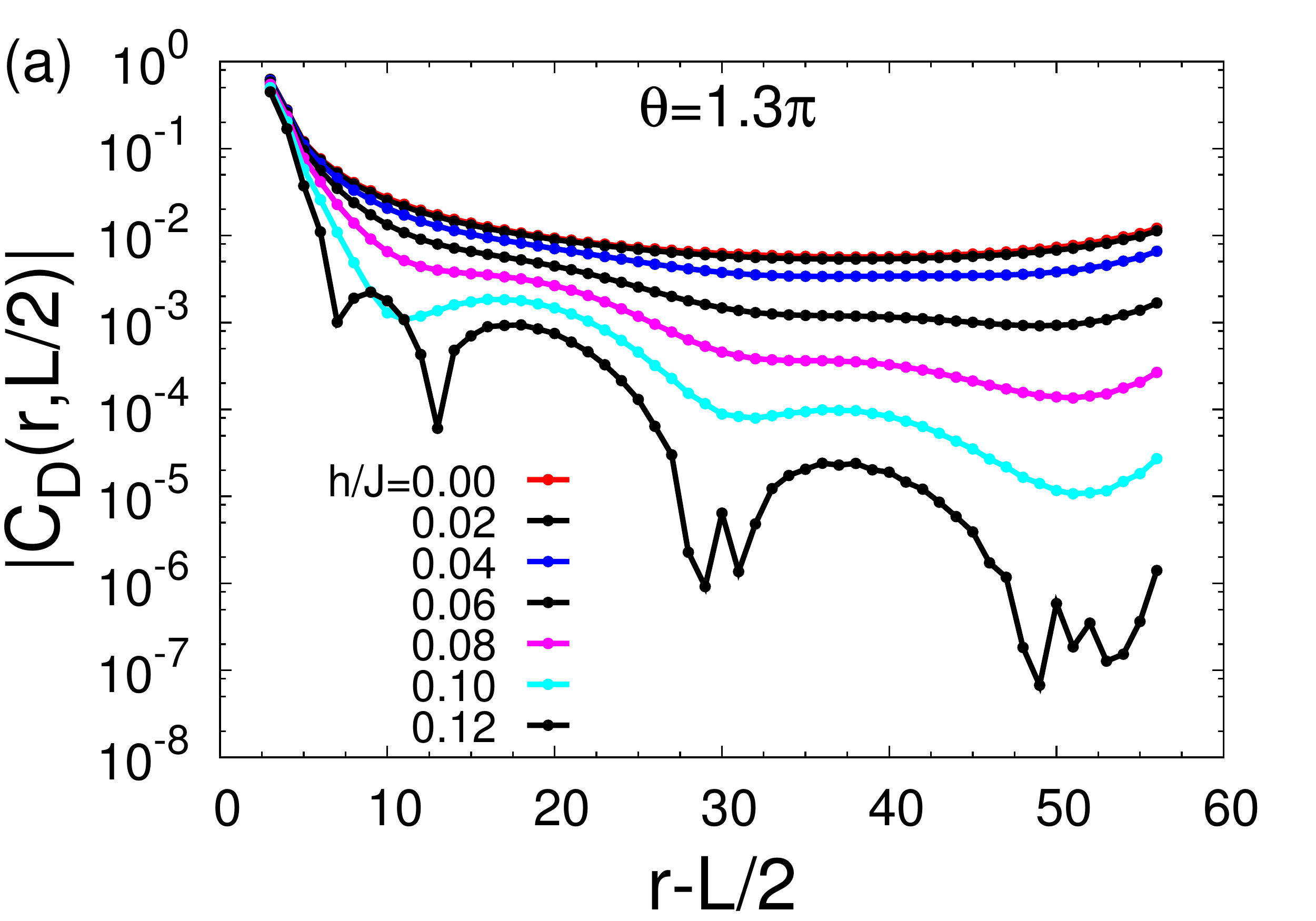}
\end{minipage}
%\newline
\begin{minipage}{.44\textwidth}
  \centering
  \includegraphics[width=0.7\linewidth,angle=-90]{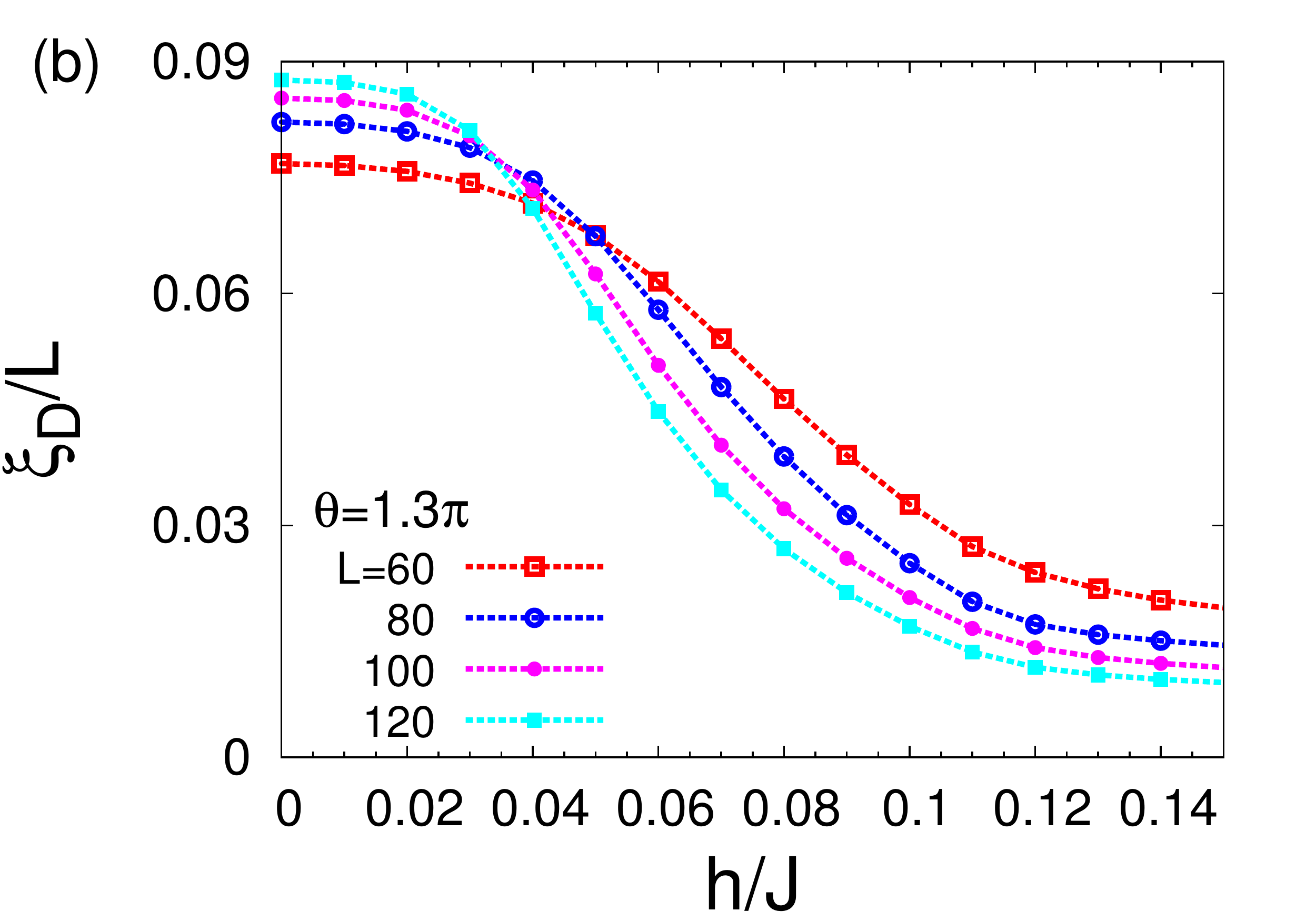}
\end{minipage}
\begin{minipage}{.44\textwidth}
  \centering
  \includegraphics[width=0.7\linewidth,angle=-90]{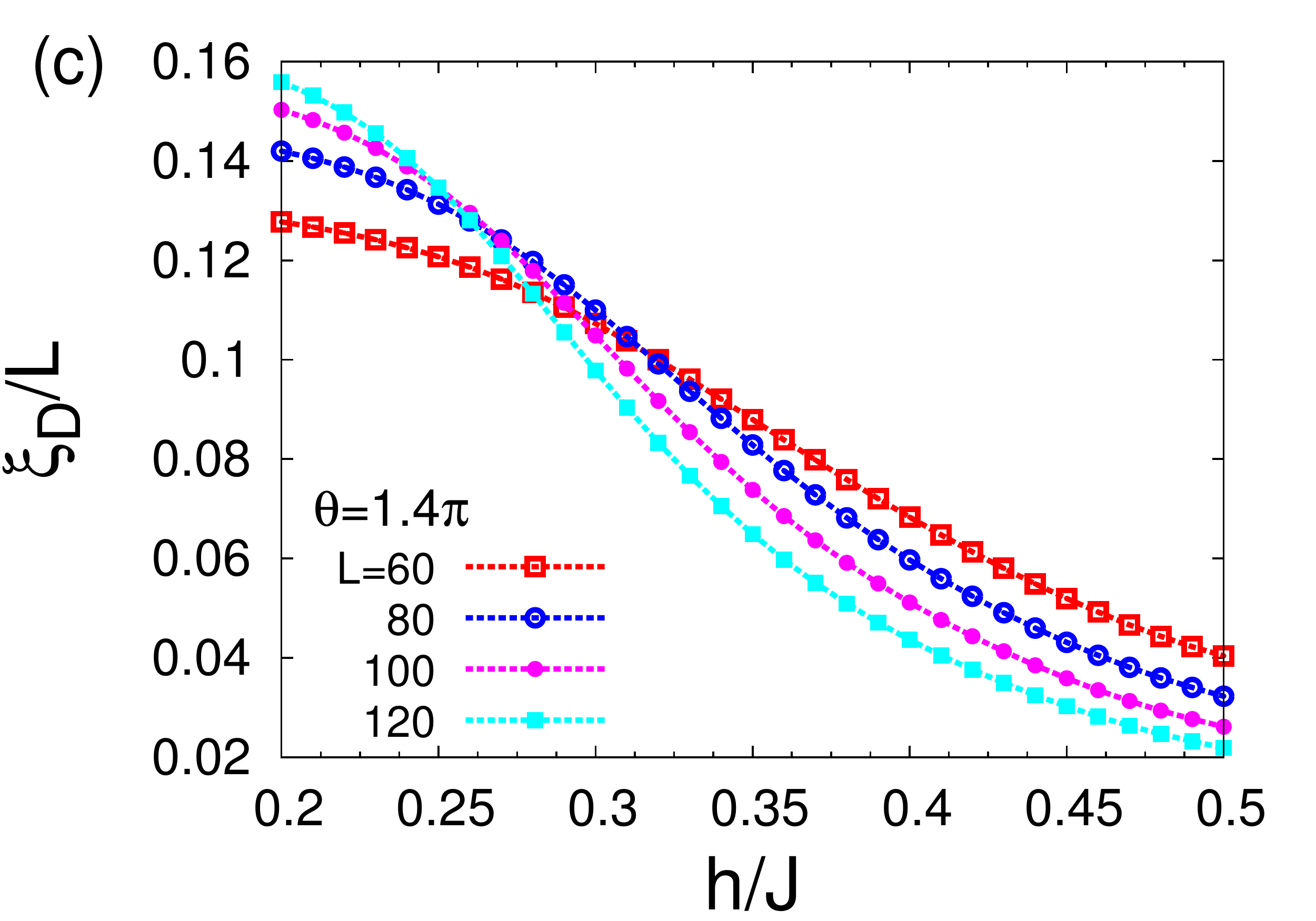}
\end{minipage}
\caption{Dimer correlations for $\theta=1.3\pi$ and $M=225$. (a) Dimer correlation function measured from the center of the chain ($r'=L/2$) for various values of the spiral magnetic field and a system size $L=120$. For small fields in the SD phase $C_D(r,L/2)$ saturates to a constant at at large $r$, where as in the SPM phase it goes to zero. The dimer correlation length [from $C_D(r,L/2)$ in Eq.~(\ref{eqn:xi})] as a function of the magnetic field for various system sizes for $\theta=1.3\pi$ (b) and $\theta=1.4\pi$ (c). The crossing of $\xi_D/L$ estimates the location of the quantum phase transition into the SPM phase, here we find $h_c(\theta=1.3\pi)/J=0.033(8)$ and $h_c(\theta=1.4\pi)/J=0.25(2)$. 
The drift in the crossing with $L$ implies large corrections to finite size scaling.} 
\label{fig:CDr}
\end{figure}

To study the long range dimer order in more depth we compute the dimer-dimer correlation function, defined as
\begin{equation}
C_D(r,r') = \langle d(r) d(r') \rangle.
\end{equation}
It is helpful to note that in the limit of $h=0$ and $\theta=1.5\pi$, the result $C_D(r,0) \sim (-1)^r c_0$ at large $r$ is exact~\cite{Sorensen-1990} and $c_0$ can be taken as another order parameter for the dimer phase. As shown in Fig.~\ref{fig:CDr} (a), we find that for weak magnetic fields, $C_D(r,L/2)$ saturates to a constant at large $r$, whereas at large fields $c_0$ goes to zero (passing over several orders of magnitude in our DMRG data). To determine the phase boundary separating the SD and SPM phases we compute the dimer-dimer correlation length $\xi_D$ from Eq.~(\ref{eqn:xi}) (replacing $C^z$ with $C_D$). This is shown in Figs.~\ref{fig:CDr} (b) and (c) for various system sizes. We find a crossing of $\xi_D/L$ vs $h/J$ for various $L$. However, in comparison with the quality of the data for $\xi^z$ in Fig.~\ref{fig:xiL}, we find that the crossing in $\xi_D$ is \emph{drifting with $L$}.
This implies that there are large finite size corrections at these system sizes, and we cannot provide as precise an estimate of $h_c$. As a result, at these system sizes we cannot accurately compute the critical exponents governing the transition for the SD to SPM quantum phase transition. This is in contrast to the SFM to SPM transition where we can reach much larger system sizes due to the small bond dimension, and thus we can estimate $h_c$ and the corresponding critical exponents accurately. Nonetheless, we use this procedure to provide an estimate of the location of the phase boundary separating the SD and SPM phases in the range $1.29 \le \theta \le 3\pi/2$.

\section{Phase Diagram}
\label{sec:pd}

\begin{figure}[t]
\centering
  \includegraphics[width=0.85\linewidth]{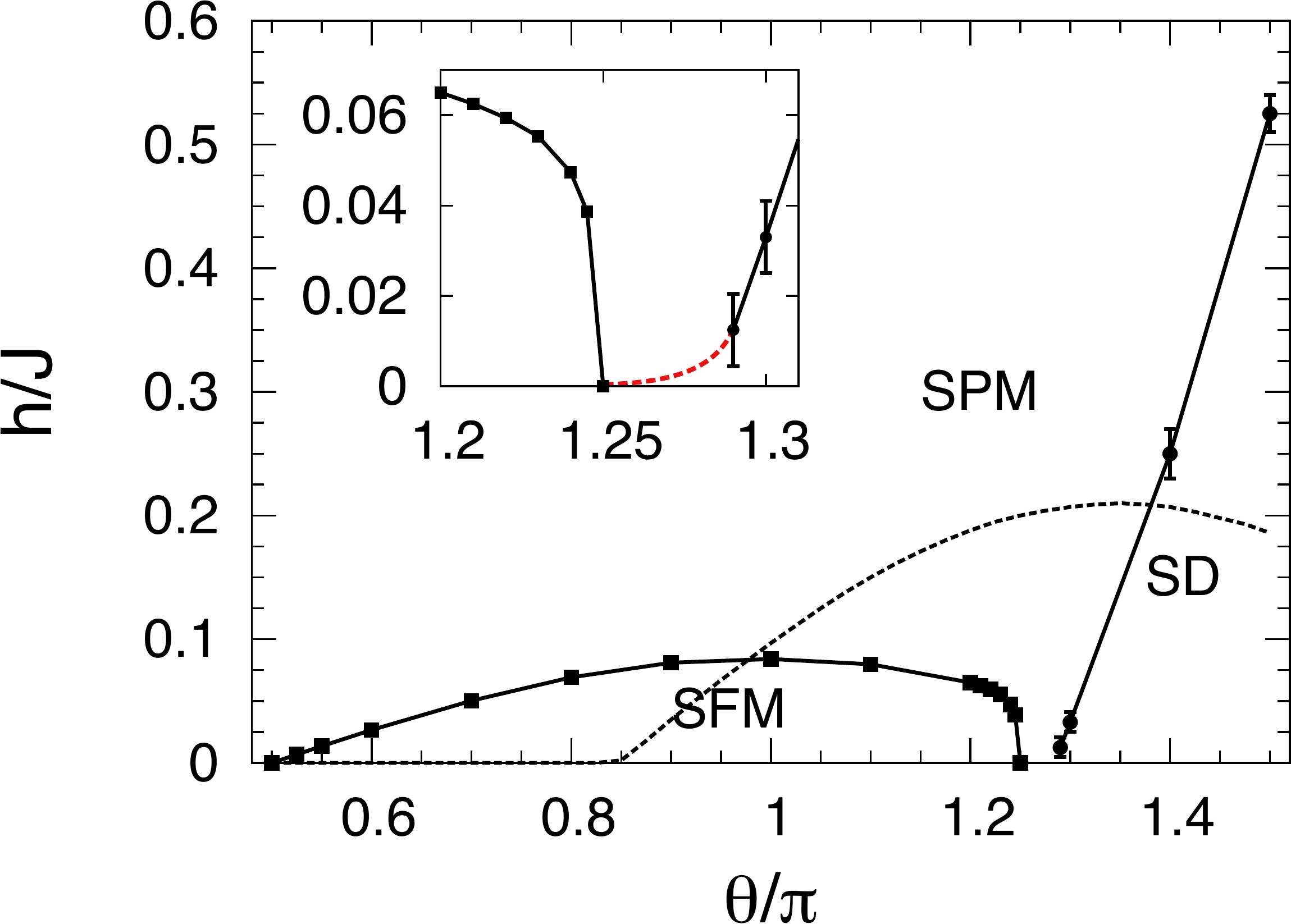}
\caption{Phase diagram as  a function of $h$ and $\theta$ for a fixed pitch $\eta=0.1\pi$ extracted from the DMRG calculations. The dashed line marks the classical SFM phase boundary, which completely misses the dimerized phase and underestimates the stability of the SFM for positive $\tilde{K}$. Error bars are determined based on the spread of the crossing in the corresponding correlation length (the error bars for SFM-to-SPM are smaller than the symbol).  (Inset) Zoomed in on the region near $\theta=1.25\pi$, here we have added a schematic dashed red line, where we expect the critical field is exponentially small.} 
\label{fig:phase_diagram}
\end{figure}

We are now in an excellent position to construct the full phase diagram of the model in the $h-\theta$ plane, restricted to the FM regime $\pi/2 < \theta < 3\pi/2$. As we have established in Sec.~\ref{sec:sf}, for $0.5\pi < \theta \lesssim 1.25 \pi$ the SFM phase undergoes a second order quantum phase transition into a SPM phase as a function of the strength of the magnetic field. In contrast, for $1.25\pi \lesssim \theta \le 1.5 \pi$ we find the model is in the SD phase that spontaneously breaks the translational symmetry for weak magnetic fields. Increasing the magnetic field ``melts'' this dimer order giving rise to a continuous transition into the SPM phase. This results in the phase diagram shown in Fig.~\ref{fig:phase_diagram}, where we have also included the classical SFM phase boundary.

\begin{figure}[t]
\centering
\begin{minipage}{.225\textwidth}
  \includegraphics[width=0.75\linewidth,angle=-90]{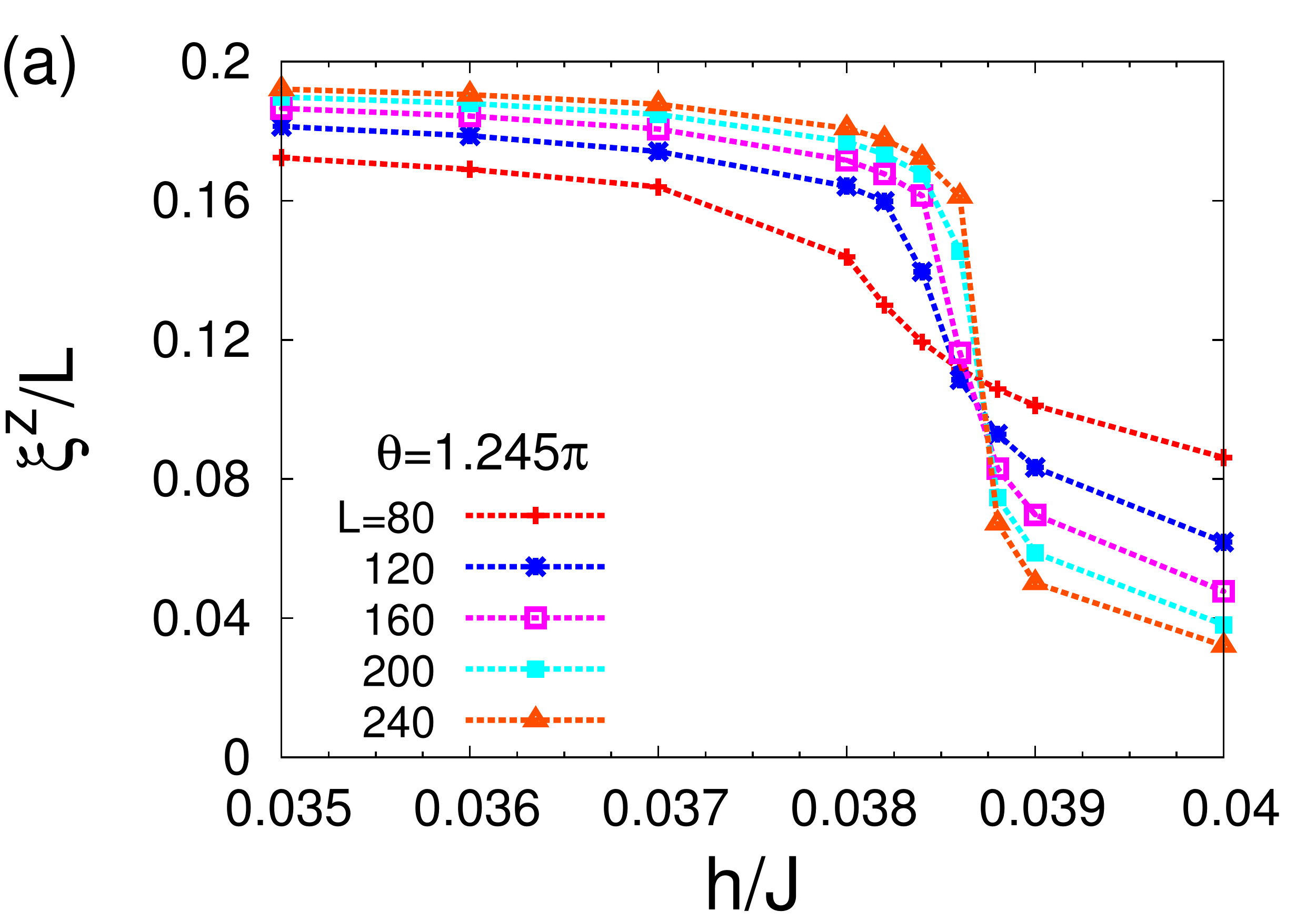}
  \end{minipage}
  \begin{minipage}{.225\textwidth}
  \includegraphics[width=0.75\linewidth,angle=-90]{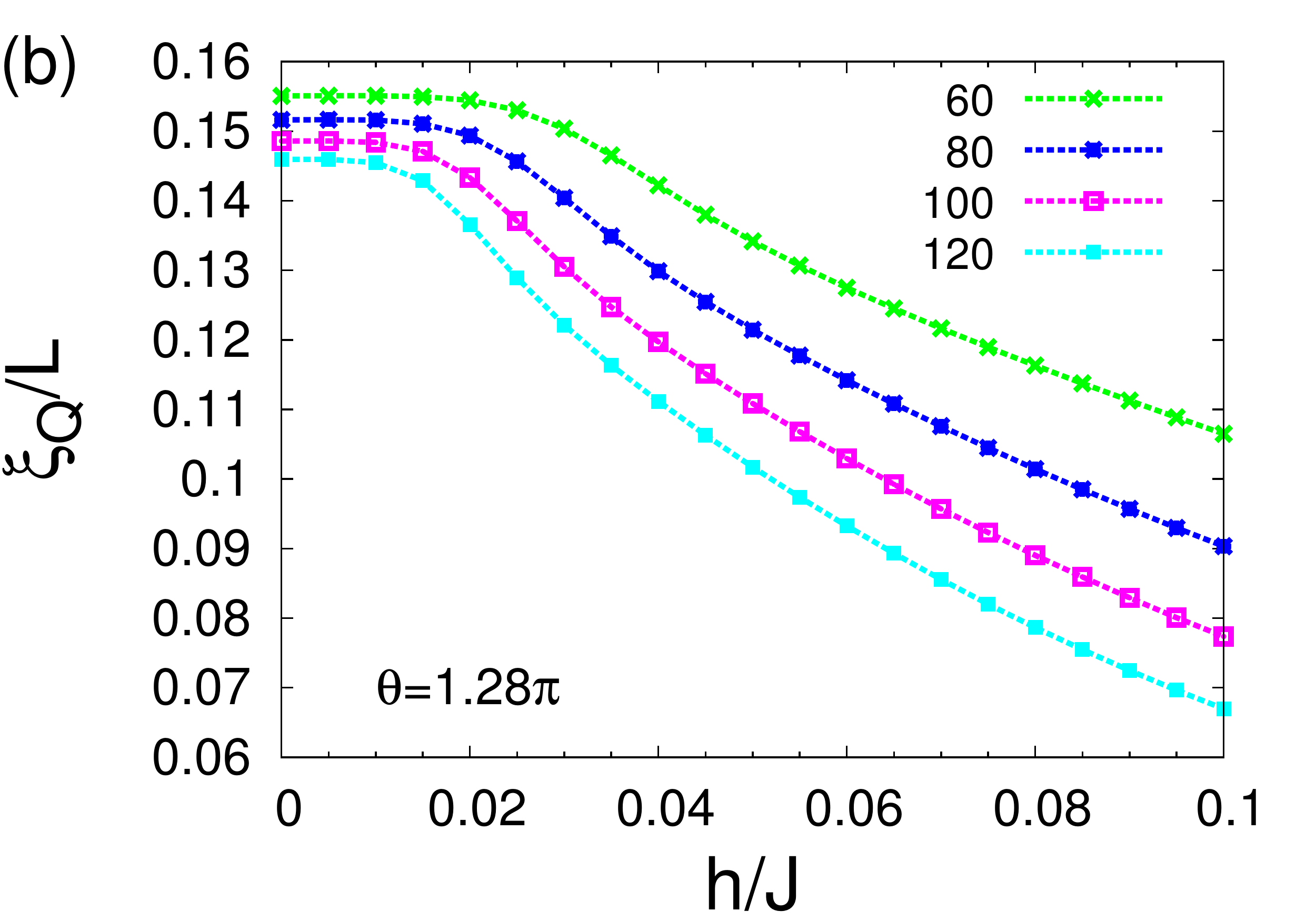}
  \end{minipage}
\caption{(a) Spin correlation length as a function of $h$ for $\theta=1.245\pi$. In close proximity to $\theta=1.25\pi$ we find the SFM to SPM transition remains continuous. (b) Quadrupolar correlation length for $\theta=1.28 \pi$, as a function of $h$ for various systems sizes. We find $\xi_Q$ is going to zero with $L$ and monotonically decreases with increasing $h$. Thus, the spiral magnetic field suppresses the quadrupolar correlations.} 
\label{fig:th1245}
\end{figure}

For $\theta$ in close proximity to $\theta=1.25\pi$ the problem is rather delicate. To study the evolution of the $\theta$ tuned transition at $\theta=1.25\pi$ in the presence of a SOC requires a full solution of the model in the three dimensional parameter space of $\theta-h-\eta$. Here, however, we have fixed $\eta$ and studied the evolution of the model as a function of $h$ and $\theta$. Coming from the SFM side of the transition ($\theta < 1.25 \pi$), for fixed $\theta=1.245\pi$ we find the crossing in 
$\xi^z/L$ is sharpening up with increasing $L$, see Fig.~\ref{fig:th1245} (a) compared with Fig.~\ref{fig:xiL} (a) and (c). Despite this, we expect the SFM to SPM transition remains continuous all the way to $\theta=1.25\pi$, where our data is consistent with no SFM all the way down to $h=0$. 

In the absence of a SOC, it was suspected that there was an intervening nematic phase separating the FM and dimer phases in the range $1.25\pi < \theta < 1.33 \pi$ (see Refs.~\onlinecite{Chubukov-1990,Chubukov-1991,Matteo-2005,Lauchli-2006,Hu-2014} and many more references therein).
It is possible, though not likely, that the spiral magnetic field can help stabilize the possible nematic phase.  
The nematic phase is expected to have a non-zero quadrupolar moment, which can be captured by the quadrupole correlation function averaged over the solid angle~\cite{Gerster-2014}. This is defined as
\begin{eqnarray}
C_Q(r)&=&\frac{2}{15}\sum_{\alpha} \langle (S^{\alpha}_n)^2(S^{\alpha}_{n+r})^2 \rangle - \langle (S^{\alpha}_n)^2 \rangle  \langle (S^{\alpha}_{n+r})^2 \rangle
\nonumber
\\
&+& \frac{1}{15}\sum_{\alpha<\beta} \langle T^{\alpha\beta}_nT^{\alpha\beta}_{n+r} \rangle - \langle T^{\alpha\beta}_n \rangle   \langle T^{\alpha\beta}_{n+r} \rangle, 
\label{eqn:Q}
\end{eqnarray}
where $T^{\alpha\beta}_n=\{S^{\alpha}_n,S^{\beta}_n \}$ is the anticommutator and we take $n=L/2$.
However, the quality of agreement between recent arguments for non-perturbative Berry phase effects and numerical data seem to rule out the nematic phase~\cite{Hu-2014,Gerster-2014} for $h=0$. Instead, the dimer order parameter is exponentially small but non-zero near $\theta=1.25\pi$ and the quadrupolar correlation length $\zeta_Q$, defined from $C_Q(r)\sim \exp(-r/\zeta_Q)/r^{x}$, is exponentially large but finite. 
Therefore, the situation is numerically very delicate already at $h=0$ 
and only very large system sizes~\cite{Hu-2014} can sort out the situation. 
Since in the presence of SOC the mentioned absence of good quantum numbers
prevents us from boosting the simulations to reach such lengths, 
we have preferred to avoid estimates of the dimer order parameter for $1.25 \pi < \theta < 1.29 \pi$.
Conversely, in this region we have extracted the effective quadrupolar correlation length $\xi_Q$
just in a similar fashion as $\xi^{z}$, by replacing $C^z$ in Eq.~(\ref{eqn:xi}) by $C_Q$ defined in Eq.~(\ref{eqn:Q}).
As shown in Fig.~\ref{fig:th1245} (b), we find that $\xi_Q / L$ decreases with increasing $L$ and $h$,
thus clearly signaling a suppression of the quadrupolar correlations by the spiral magnetic field.

We are now in a position to complete the phase diagram.
Since we have established that dimerization is stable in the presence of SOC and the nematic phase is not, we argue that in the regime $1.25 \pi \le \theta < 1.29\pi$ of the phase diagram, the SD phase should be stable for an exponentially small field strength and the phase boundary is beyond the scope of our present numerics. To show this we have placed a schematic red line in the phase diagram.

\section{Discussion and Conclusion}
\label{sec:dc}
We have described the evolution of the ferromagnetic spin-1 bilinear-biquadratic spin chain, an effective description of the Mott insulating phase of a spin-1 Bose-Hubbard model, in the presence of Raman-induced spin-orbit coupling. Our high-precision numerical results, obtained by (essentially exact) density matrix renormalization group calculations, have revealed a rich magnetic phase diagram, with phases that ``survive'' the application of a spiral magnetic field, and unusual quantum phase transitions to the high-field paramagnetic state. If we work in the co-rotating frame of the Raman field, the spiral magnetic field transforms into translation-invariant spin-anisotropic and antisymmetric Dzyaloshinskii-Moriya interaction terms (see Eq.~\ref{eqn:dmint}), which have been rarely  studied for spin-1 degrees of freedom. 
Our work in this regard is unique. 

Our calculated
critical exponents for the spiral ferromagnet to paramagnet quantum phase transition in the spin orbit coupled spin-1 Bose-Hubbard model ($\nu \approx 2/3$ and $\gamma\approx 1/2$) 
point to the possible existence of a new universality class in this problem. Our results for the numerical value of the correlation length exponent
 are consistent with the universality class of classical uniaxial dipolar ferromangets~\cite{Larkin-1969,ZinJustin,Aharony-1973} and the XY chiral spin liquid transition in two dimensions~\cite{Dimitrova-2014} but the current numerical accuracy of our calculations does not allow us to disentangle this from the possible three-dimensional Ising and XY universality classes. On physical grounds however, it is most natural to expect that our results are consistent with the chiral spin liquid class as this involves a \emph{helical} magnetic transition.  It will be very interesting to study the dynamical properties of this universality class in more depth to determine the value of the dynamic exponent $z$. In the absence of a spin orbit coupling the quantum ferromagnet has spin waves dispersing like $q^2$ and therefore lives in an effective dimension $d_{\mathrm{eff}} = d+z = 3$. However,  it is not clear from our study of the static quantum critical properties what the value of $z$ is in the presence of a spin orbit coupling. If this transition is consistent with the chiral class in classical $d_{\mathrm{eff}} = 2$, it would imply that the presence of the spin orbit coupling and the corresponding spiral magnetic field has induced a \emph{dimensional reduction} (i.e. from 3 to 2 and correspondingly $z$ goes from 2 to 1) at the spiral ferromagnet to paramagnet transition.  
% This, if valid, is a most remarkable finding since naively one expects quantum systems to have an additional dimension (i.e. imaginary time) compared with the corresponding classical systems for critical phenomena (i.e. dimension enhancement rather than reduction).
 We expect the universality class we have discovered to be prevalent in spin orbit coupled spinor bosons in one dimension as our results  describe a ferromagnet in a spiral magnetic field. Thus we expect that our results will also describe the universality class of spin orbit coupled pseudospin-1/2 bosons in the insulating Mott phase. 
We expect that other unusual universality classes in spin-1 chains 
can be realized with the addition of 
long-range interactions, which can be emulated in trapped ion simulators~\cite{Senko-2015} and polar molecules~\cite{Kestner-2011}. 

An inherent feature of the spin-1 nature of the model is the existence of the spiral dimer phase, which will not be present in the pseudospin-1/2 Bose Hubbard model (without fine tuning). 
It is then natural to ask: what is the nature of the spiral \emph{dimer} to paramagnet quantum phase transition? It is not yet clear whether or not the transition can simply be described by an Ising transition due to the $Z_2$ dimer order parameter, or if spin-orbit coupling has changed the problem fundamentally as it has for the spiral ferromagnet to paramagnet transition. To address this question, it will be essential to reach large chain lengths and large bond dimensions in the presence of spin orbit coupling. It will be interesting to try and generalize the field theory of the dimer phase~\cite{Hu-2014} to incorporate a spin orbit coupling.

Our results provide  guidance for future cold atom experiments to try and probe the exotic magnetic phases and the corresponding phase transitions we have discovered here theoretically. 
We can place each spin-1 bosonic atom that is readily trapped and cooled on the phase diagram in Fig.~\ref{fig:phase_diagram} for the unit filled Mott lobe. It is possible to study the spiral dimer phase by using $^{23}$Na with $\theta = 1.26 \pi$. Whereas, the spiral ferromagnet is accessible
to $^7$Li, $^{41}K$, and $^{87}$Rb with $\theta=1.15 \pi, 1.242\pi$ and $1.249\pi$ respectively. It will be very promising to probe the ferromagnetic spiral phase dynamically in experiments by preparing a polarized (i.e. ferromagnetic) initial state, which quenches the spin-entropy. Then adiabatically or diabatically tuning the system into the Mott phase, should allow experiments to probe the physics of the spiral ferromagnetic ground state. We believe that the cold atom realization of the one dimensional bosonic spin system discussed in our work will lead to the observation of several new strongly correlated quantum magnetic phases which do not exist in solid state materials.

\acknowledgements{We would like to thank Pallab Goswami for useful discussions and Stefan Natu for collaborations at the early stages of this work. J.H.P. and I.B.S. thank Hilary Hurst and Justin Wilson for collaborations on related work. 
This work is partially supported by JQI-NSF-PFC, LPS-MPO-CMTC, and Microsoft Q (J.H.P., W.S.C., and S.D.S.), the Physics Frontier Center seed grant ÒEmergent phenomena in interacting spin-orbit coupled gasesÓ (J.H.P. and W.S.C), AFOSRs Quantum Matter MURI, NIST, and the NSF through the PFC at the JQI (I.B.S.). M.R. acknowledges the hospitality of the Condensed Matter Theory Center at the University of Maryland.
The calculations where performed using the open-source code ``Powder with POWER'' (or dmrg.it) using the University of Maryland supercomputing resources (http://www.it.umd.edu/hpcc) and
the MOGON Cluster at the JGU Mainz, made available by the AHRP initiative
 in conducting the research reported in this paper.}

\bibliography{s1BH-SOC}

\end{document}